\documentstyle[A4,12pt]{article}
\newcommand{\nc}{\newcommand}
\nc{\rt}[1]{\mbox{$\tilde{\rho}_{#1}$}}
\nc{\mt}[1]{\mbox{$\tilde{\mu}_{#1}$}}
\nc{\ct}[1]{\mbox{$\tilde{c}_{#1}$}}
\nc{\mmu}[1]{\mbox{${\mu}_{#1}$}}
\nc{\ro}[1]{\mbox{${\rho}_{#1}$}}
\nc{\D}[1]{\mbox{$D^{#1}$}}
\nc{\alf}[1]{\mbox{$\alpha_{#1}$}}
\nc{\bet}[1]{\mbox{$\beta_{#1}$}}
\nc{\ag}[1]{\mbox{$a_{#1}$}}
\nc{\bg}[1]{\mbox{$b_{#1}$}}
\nc{\rat}{\textstyle\frac}
\nc{\Llra}{\Longleftrightarrow}
\nc{\llra}{\longleftrightarrow}
\nc{\ra}{\rightarrow}
\nc{\osp}{$osp(1\!\mid\!2)$}
\nc{\sln}{$sl(n+1\!\mid\!n)$}
\nc{\f}[1]{\frac{#1}{2}}
\nc{\om}{\Omega}
\nc{\omt}{\tilde{\Omega}}
\nc{\os}{\Omega^{\ast}}
\nc{\Cbar}{\mbox{C}\!\!\!\mbox{I}\,\,}
\nc{\oms}{\Omega^{\ast}}
\nc{\omst}{\tilde{\om}^{\ast}}
\nc{\Ga}{\gamma_{1}}
\nc{\Gb}{\gamma_{2}}
\nc{\rha}{\tilde{\rho_{3}}}
\nc{\rhb}{\tilde{\rho_{2}}}
\nc{\drp}{\partial}
\nc{\mua}{\tilde{\mu_{3}}}
\nc{\mub}{\tilde{\mu_{2}}}
\nc{\epc}{\epsilon_{c}}
\nc{\epa}{\epsilon_{1}}
\nc{\epb}{\epsilon_{2}}
\nc{\epi}{\epsilon_{i}}
\nc{\E}{\Upsilon}
\nc{\Ea}{\Upsilon_{1}}
\nc{\Eb}{\Upsilon_{2}}
\nc{\Th}{\Theta}
\nc{\Db}{\bar{D}}
\nc{\gch}{grassmannian character }
\nc{\gdg}{grassmannian degree }
\nc{\zab}{z_{12}}
\nc{\teab}{\theta_{12}}
\nc{\tazab}{\frac{\teab}{\zab}}
\nc{\tbar}{\bar{\theta}}
\nc{\spc}{{\cal R}}
\nc{\ca}{\mbox{\boldmath $A$}}
\nc{\cat}{\tilde{\ca}}
\nc{\sz}{\mbox{\boldmath $z$}}
\nc{\szb}{\bar{\sz}}
\nc{\con}{{\cal A}}
\nc{\cont}{\tilde{\con}}
\nc{\db}{\bar{\partial}}
\nc{\dbar}{\bar{\partial}}
\nc{\dh}{{\cal D}}
\nc{\R}{\mbox{I}\! \mbox{R}}
\nc{\dlbd}{\mbox{$d\lambda$}}
\nc{\dlbdb}{\mbox{$d\bar{\lambda}$}}
\nc{\Ct}{\tilde{C}}
\nc{\sig}[2]{\mbox{$\sigma_{#1}^{#2}$}}
\nc{\bel}{H^{\ z}_{\bar{\theta}}}
\nc{\bol}[2]{\mbox{${\cal L}_{#1}(#2)$}}
\nc{\beqna}{\begin{eqnarray}}
\nc{\eeqna}{\end{eqnarray}}
\nc{\beqnas}{\begin{eqnarray*}}
\nc{\eeqnas}{\end{eqnarray*}}
\nc{\zbar}{\bar{z}}
\nc{\poi}{PB\ }
\nc{\dta}{\delta^{(2)}(\sz_{1}-\sz_{2})}
\nc{\epo}{\epsilon_{0}}
\nc{\epd}{\epsilon_{d}}
\nc{\epe}{\epsilon_{e}}
\nc{\uz}{\Upsilon^{Z}}
\nc{\ut}{\Upsilon^{\Th}}
\nc{\lzz}{\Lambda_{z}^{\ Z}}
\nc{\unit}{{1\:\!\!\!\mbox{\small{l}}}}
\nc{\sv}{super-Virasoro}

\nc{\pl}[3]{Phys. Lett. {\bf B#1} (#2) #3}
\nc{\np}[3]{Nucl. Phys. {\bf B#1} (#2) #3}
\nc{\cmp}[3]{Commun. Math. Phys. {\bf #1} (#2) #3}
\nc{\jmp}[3]{J. Math. Phys. {\bf #1} (#2)#3}
\nc{\ijmp}[3]{Int. J. Mod. Phys. {\bf A#1} (#2) #3}
\nc{\mpl}[3]{Mod. Phys. Lett. {\bf  A#1} (#2) #3}
\nc{\cqg}[3]{Class. Quant. Grav. {\bf #1} (#2) #3}
\nc{\rep}[3]{Phys. Rep. {\bf #1} (#2) #3}

\begin{document}

\thispagestyle{empty}
\begin{flushright}
LPTB/94-19
\end{flushright}

\vspace*{15mm}
\begin{center}
{\bf \Huge{A geometrical approach to}}
\end{center}
\begin{center}
{\bf \Huge{ super $W$-induced gravities }}
\end{center}
\begin{center}
{\bf \Huge{in two dimensions}}
\end{center}
\bigskip
\centerline{\bf Jean-Pierre Ader, Franck Biet, Yves Noirot}
\vspace{8mm}
\centerline{\it Centre de Physique Th\'eorique et de Mod\'elisation
 de Bordeaux}
\centerline{CNRS $ ^{{\bf \dag}}$, {\it Universit\'e de Bordeaux I}}
\centerline{\it 19, rue du Solarium}
\centerline{\it F - 33175 - Gradignan Cedex}
\vspace{10mm}

\begin{abstract}
A geometrical study of supergravity defined on $(1\!\mid\!1)$
complex superspace is
presented. This approach is based on the introduction of generalized
superprojective structures extending the notions of super Riemann geometry
to a kind of super W-Riemann surfaces. On these surfaces a connection is
constructed. The zero curvature condition leads to the super Ward
identities  of the underlying supergravity. This is accomplished through
the symplectic form linked to the (super)symplectic manifold of all super
gauge connections. The BRST algebra is also
derived from the knowledge of the super W-symmetries which are the gauge
transformations of the vector bundle canonically associated to the
generalized superprojective structures. We obtain
the possible consistent BRST (super)anomalies
and their cocycles related by the descent equations.
Finally we apply our considerations to the case of supergravity.
\end{abstract}

\nopagebreak
\vspace{15mm}
\begin{flushleft}
\rule{2in}{0.03cm} \\

{\footnotesize \ ${}^{\dag}$
Unit\'e Associ\'ee au CNRS, U.A. 764.} \\ [-0.04cm]
\vskip 0.07truecm
\end{flushleft}

\section{Introduction}
In 1985 Zamolodchikov introduced new symmetries in conformal models generated
by currents of spin higher than 2, whose commutation relations were shown to
have non linear terms. These new algebras, the $W$-algebras, were also shown
to
appear in integrable systems through Poisson brackets. They lead to induced
(classical) $W$-gravities which are higher-spin gauge theories in two
dimensions whose gauge algebras are these $W$-algebras just as the Virasoro
algebra appears as the residual symmetry of gauge fixed gravity in two
dimensions.
 For recent reviews see \cite{r1,r2}. It is natural to interpret such
 a class of conformal field theories as possible realizations of the
$W$-geometries introduced in \cite{SSZ,zucchini,gerv}. The approach of
\cite{SSZ,gerv} starts with the embedding of a 2-dim base manifold into a
$n-1$ dimensional K\"ahler manifold whereas \cite{zucchini} is a recent
development in the light-cone gauge. In this paper we generalize to the
(1,1) supersymmetric case the geometrical setting given in
\cite{zucchini}.

     The numerous links between 2D gravity and integrable systems through
Poisson brackets are well known. Since the evolution of an integrable system
can be thought of as a zero-curvature condition associated with some gauge
group it seems natural to consider theories of the $W_n$ induced gravities
based on such a condition. The starting point of these theories is the
vanishing condition of the field strength associated to a pair of matrices
($A_z, J_{\bar{z}}$), giving in the standard case the chiral Virasoro
 Ward-identity. After all, since the success of the Polyakov formulation
\cite{Pol}
was to show that the unexpected $SL(2,\R)$ current algebra arises in 2D
gravity in the light-cone gauge, it is not astonishing that the first attempts
use a group approach, the generalization to higher $W$-gravities consisting
of replacing $SL(2,\R)$ by some other non-compact real Lie group.
 More precisely the current $J_{\bar{z}}$ is
parametrized as $J_{\bar{z}}=h^{-1} \bar{\partial} h$  whereas
$A_z=g^{-1} \partial g$ where $h,g$ are some group valued functions
\cite{pw}. The matrix $A_z$ contains the projective connection and the
fields associated to it.

   More recently Zucchini \cite{zucchini} has presented a formalism in which
the
usual Riemann surface is embedded in a $n$-dimensional complex manifold to
which is canonically associated a $SL(n,\Cbar)$ fiber bundle.
On this bundle
a connection $\con$ with zero curvature is defined.
 This connection appears as a pair of matrices ($\om,\oms$) which can be
parametrized by introducing projective structures ($\mu_i,\rho_i$)
generalizing the
pair ($\mu^{\ z}_{\bar{z}},\rho_{zz}$) consisting of the Beltrami coefficient
and the projective connection respectively on the ordinary Riemann surface.
Working in this framework one finds $(n-1)$ pairs of generalized Beltrami
differentials and projective connections characterizing a kind of ``$W_n$''
Riemann surface, which is assumed to be the geometrical way to reach the
basic notions behind the $W_n$ gravity theories i.e. the
$W_n$-algebras. The geometric structure underlying these algebras appears as
extra data on the Riemann surface.
The zero curvature condition  on the connection $\con$ is
naturally ensured by the definition of its components in terms of a basic
matrix $W$, namely
$\om \equiv \drp W W^{-1}\;,\;\oms \equiv \bar{\drp} W W^{-1}$.
 In the standard cases $(W_2, W_3)$ the expression of the resulting matrices
($\om,\oms$) as functions of the gauge fields and the spin-s currents,
respectively, is identical
to the result of the group  theory approach \cite{OSSV} when $h \equiv g$
(the connection
$\con$ being a pure gauge). Moreover the local expression of $h$ can be
obtained in that case by taking a Gauss decomposition for $h$ and the
two formalisms coincide since $W=h^{-1}$, Zucchini's formalism
providing a geometrical interpretation for the physical fields
entering in the $W$-gravities.

The essential advantage of this approach is to define the $W_n$ symmetries as
 gauge transformations of the flat vector bundle canonically associated to
the generalized projective structures. From the knowledge of these symmetries
 the off-shell nilpotent BRST algebra for an arbitrary $W_n$ model is derived
 \cite{zucchini}. Several other advantages can
be emphasized: the generalization to arbitrary $W_n$ models is automatic and
is uniquely limited by technical complications; gluing properties of the
fields under conformal coordinate changes are known ``ab initio'' and result
from the formalism itself. Furthermore this
formulation allows us to interpolate between various $W$ theories thus taking
into account their ``nested'' structure, $W_n \subset W_{n+1}$, where the
inclusion symbol indicates that the formulation of $W_n$ can be obtained
from $W_{n+1}$ by setting to zero the projective variables occuring at
the level $n+1$.

  Most of the existing results, with some exceptions \cite{theisen}, concern
the bosonic theory; complete studies in the supersymmetric case are still
lacking. Accordingly, the systematic manifestly (1,1) supersymmetric
extension of \cite{zucchini} presented here is an attempt to fill this gap.
  First we show (sect.2) that generalized superprojective
structures may be parametrized
in a one-to-one fashion by pairs of superfields which generalize the
super-Beltrami differential and the superprojective connection.
In supersymmetry, besides obvious technical difficulties, one has to face
features which do not appear at the bosonic level. In particular, since some
superfields involve only non physical fields, it seems natural to restrict
the geometry by turning them off. When going to higher $n$ the number of
possibilities of this kind increases and the full model, although very
cumbersome, can give birth to several meaningful and interesting
developments.

    The physical geometrical fields are not the fields ($\mu_i,\rho_i$) which
emerge naturally from the construction since in general these objects do not
change in a homogeneous way under conformal coordinate
transformations. In fact
the physical fields $\Phi \equiv (\tilde{\mu_i},\tilde{\rho_i})$ are
sections~\footnote{except q=2 which concerns the particular case of the
projective connection.}
($\bar{k}k^{-p}$ with $1\leq p\leq n-1$ and $k^q$ with $3\leq q\leq n$)
of the fiber bundle which transform as differentials:
$\Phi_b=\Phi_a(k)^h(\bar{k})^{\bar{h}}$, $h$ and $\bar{h}$ being the
conformal weights of $\Phi$. They appear as combinations of the
($\mu_i,\rho_i$). A method exists to construct systematically the
$\tilde{\rho_i}$ using the approach of \cite{ds}. Having obtained these
fields by transposing these ideas to the supersymmetric case, we use
the natural connection $\con$ to construct a symplectic form
which provides
a systematic way of obtaining the $\tilde{\mu_i}$ (sect.3).

  In sect.4, the BRST off-shell nilpotent algebra associated to an arbitrary
 super $W_n$ induced gravity \footnote{Here the index $n$ is choosen with
reference to the underlying $W_n$ model which is the bosonic limit of the
super model considered.}
($SW_n$), is constructed from the $SW_n$ symmetries which are
defined as gauge transformations of the vector bundle canonically associated
to the	generalized superprojective structures. We present a general
formulation of a consistent and covariant (i.e.well-defined on the
super Riemann surface and obeying the Wess-Zumino
consistency condition) super anomaly which may occur
and of the cocycles linked to it by the BRST operator.
In the last section we discuss the example of the induced supergravity
($SW_2$), writing the model in its full generality and comparing our results
 with existing ones.

 For sake of clarity some details are collected in an appendix.

\section{Geometrical setting}

	Starting from a $(1\!\mid\!1)$ complex superspace with coordinates
($z,\theta$) we consider a supermanifold ${\cal M}$ which is obtained by
patching together local coordinate charts
$\{V_{a},(z,\bar{z},\theta,\bar{\theta})_{a}\}$. The
basis of the tangent space is
$(\partial_{z},\partial_{\bar{z}},D,\bar{D})$ where
the super-derivatives are defined by
$$ D=\partial_{\theta}+\theta\partial_{z},$$
$$ \bar{D}=\partial_{\tbar}+\tbar\partial_{\bar{z}}.$$
They obey
\[
D^2 = \partial_z \equiv \partial
\qquad
\mbox{and}
\qquad
\bar{D}^2 = \partial_{\bar{z}} \equiv \bar{\partial}.
\]
Under a change of reference structure
$(z_{a},\theta_{a})\rightarrow(z_{b},\theta_{b})$ the vector field
$D$ transforms as
$$ D_{a}=(D_{a}\theta_{b})D_{b}+(D_{a}z_{b}-\theta_{b} D_{a}\theta_{b})
\partial_{z_{b}} +(D_{a}\tbar_{b})\bar{D}_{b}
+(D_{a}\bar{z}_{b}-\tbar_{b} D_{a}\tbar_{b})\partial_{\bar{z}_{b}}.$$
The complex supermanifold thus defined becomes a $N=1$ super Riemann
surface (SRS) $S\Sigma$ if the transition functions
$z_{b}(z_{a},\bar{z}_{a},\theta_{a},\tbar_{a}),\ \theta_{b}(z_{a},\bar{z}_{a},
\theta_{a},\tbar_{a})$
(and their complex conjugates) between two local
coordinate charts $(U_{a},(z,\bar{z},\theta,\tbar)_{a})$ and
$(U_{b},(z,\bar{z},\theta,\tbar)_{b})$
satisfy the following conditions of super-conformality
\footnote{It is understood that the complex conjugate (cc) conditions are
also to be taken into account.}
$$ \bar{D}_{a}z_{b}=\bar{D}_{a}\theta_{b}=0 \qquad \mbox{and} \qquad
 D_{a}z_{b}=\theta_{b}D_{a}\theta_{b} .$$
With these conditions the super-derivative transforms homogeneously.
An atlas of superprojective coordinates on a SRS (without boundary)
$S\Sigma$
is a collection of homeomorphisms $\{(Z,\Theta)_{\alpha}\}$
\footnote{We will always use the Greek
letters for the extended superprojective atlas and the Latin letters for the
reference atlas on $S\Sigma$.}
of $S\Sigma$ into
$\Cbar^{1 \mid 1}$, locally defined on domains
\{$K_{\alpha}$\} with the gluing laws
on overlapping domains $K_{\alpha}$ and $K_{\beta}$ \cite{CR}
\begin{eqnarray}
Z_{\beta} & = & \frac{mZ_{\alpha} + p}{qZ_{\alpha} + r}
+ \Theta_{\alpha} \frac{\gamma Z_{\alpha} + \delta}
{(qZ_{\alpha} + r)^2}  \nonumber \\
\Theta_{\beta} &=& \frac{\gamma Z_{\alpha} + \delta}{qZ_{\alpha} + r}
+\Theta_{\alpha} \frac{1}{qZ_{\alpha} + r}
(1 + \rat{1}{2}\gamma \delta)
\label{Moebius}
\end{eqnarray}
where the matrix
$\left( \begin{array}{cc} m & p \\ q & r \end{array} \right)$
belongs to $SL(2,\Cbar)$ whereas $\gamma$ and $\delta$ are odd Grassmann
numbers.

Such an atlas defines a supercomplex structure on $S\Sigma$,
 or equivalently, superconformal classes of metrics which are
related to the reference structure
$(z,\theta)$ by the super-Beltrami differentials through the super-Beltrami
equations \cite{CR,dgbel,Takama}. These structures are parametrized
by two independent odd superfields $H^{\ z}_{\bar{\theta}} $ and
$H^{\ z}_{\theta}$ (and the c.c. analogues).
Since $H^{\ z}_{\theta} $ contains only auxiliary space-time fields,
studies are
in general limited to the special case $H^{\ z}_{\theta} = 0 $, a
restriction which
is equivalent to $DZ = \Theta D\Theta$. The algebra which underlies this
framework is the well-known super-Virasoro algebra.

Our approach to the super W-algebras is based on a straightforward
generalization of the notion of superprojective coordinates. It consists in
enlarging the set of these coordinates $(Z,\Theta)$ by considering
a collection of local maps
$\{(Z^1,\ldots,Z^n;\Theta^1,\ldots,\Theta^n)_{\alpha}\}$
of $S\Sigma$ into $\Cbar^{1 \mid 1}$. These variables can be gathered in the
vector

\[
 {\cal Z} = \left(\begin{array}{ccccccc}
1 &  Z^1 & \cdots & Z^n & \Theta^1 & \cdots & \Theta^n
\end{array}
\right)^{st}
\]
where $st$ is the supertranspose and the $Z^{i},\; \Theta^{i}$ which are
functions of ($z,\theta,\bar{z},\bar{\theta}$) have respectively an even
and odd grassmannian character.
We further impose the transition functions on overlapping domains
$K_{\alpha}$, $K_{\beta}$ to be

\begin{equation}
{\cal Z}_{\beta}^i=\frac{\sum_{j=0}^{2n} \Phi_{\beta \alpha}^{ij}
{\cal Z}_{\alpha}^j }
   { (\Phi^{0}_{\beta \alpha},{\cal Z}_{\alpha})_{E}}
\label{tr}
\end{equation}
where $\Phi_{\beta\alpha}$ is a constant non singular
$(2n + 1) \times (2n + 1)$ complex matrix of super-determinant 1 and
$(\Phi^{0}_{\beta\alpha},{\cal Z}_{\alpha})_{E}$ is the euclidean scalar
product between the first row (labelled $0$) of $\Phi_{\beta\alpha}$
and the vector $\cal Z$.
These transition functions can be regarded as a generalization of the
superprojective (M\"{o}bius) transformations (\ref{Moebius}).
We now define the matrix $W_{0}$ by (the dot marks the matrix product)
    \begin{equation}
	   W_{0} = {\cal D} \cdot {\cal Z }^{st}
   \label{w0}
   \end{equation}
where $\cal D$ is the vector

   \[
       {\cal D} =  \left(
			  \begin{array}{cccccccc}

 1 & \partial & \cdots	 & \partial^{n}  & D   &  D\partial   &  \cdots   &
			     D\partial^{n-1}
			  \end{array}
		  \right)^{st}.
   \]

We divide each coordinate by the superdeterminant
$\Delta$ of $W_{0}$ (assuming that the non-singularity condition
$\Delta \neq 0$ holds everywhere on $S\Sigma$), thus defining
the matrix $W$
\begin{equation}
	 W= {\cal D} \cdot ( \frac{1}{\Delta} \, {\cal Z} )^{st}.
\label{defw}
\end{equation}
The matrix we get is

\begin{equation}
 W = \left( \begin{array}{ccccccc}

     \frac{1}{\Delta} & \frac{Z^{1}}{\Delta} & \ldots &
\frac{Z^{n}}{\Delta} & \frac{\Th^{1}}{\Delta} & \ldots &
\frac{\Th^{n}}{\Delta}	\\[3mm]

   \drp (\frac{1}{\Delta})   & \partial (\frac{Z^{1}}{\Delta}) & \ldots &
\partial (\frac{Z^{n}}{\Delta}) & \partial (\frac{\Th^{1}}{\Delta}) &
\ldots & \partial (\frac{\Th^{n}}{\Delta})  \\[3mm]

     \vdots & \vdots &	& \vdots & \vdots &  & \vdots \\

     \drp^{n}(\frac{1}{\Delta}) & \drp^{n}(\frac{Z^{1}}{\Delta}) &
\ldots & \drp^{n}(\frac{Z^{n}}{\Delta}) &
	 \drp^{n}(\frac{\Th^{1}}{\Delta}) & \ldots &
\drp^{n}(\frac{\Th^{n}}{\Delta})  \\[3mm]

     D(\frac{1}{\Delta}) & D(\frac{Z^{1}}{\Delta}) & \ldots &
D(\frac{Z^{n}}{\Delta}) & D(\frac{\Th^{1}}{\Delta}) & \ldots &
D(\frac{\Th^{n}}{\Delta})
\\[3mm]

      \vdots & \vdots &  & \vdots & \vdots &  & \vdots \\

     D\drp^{n-1}(\frac{1}{\Delta}) & D\drp^{n-1}(\frac{Z^{1}}{\Delta})
& \ldots & D\drp^{n-1}(\frac{Z^{n}}{\Delta})	&
	 D\drp^{n-1}(\frac{\Th^{1}}{\Delta}) & \ldots &
D\drp^{n-1}(\frac{\Th^{n}}{\Delta})

	       \end{array}
	       \right)	
\end{equation}

We will study the transformation law of this matrix under both
an extended
superprojective transformation ${\cal Z}_{\alpha}\rightarrow{\cal Z}_{\beta}$
and a change of the local coordinates
$\sz_{a}\rightarrow \sz_{b}$ where $\sz=(z,\theta)$; it reads
     \begin{equation}
	W_{b \beta}={\cal D}_{b} \cdot ( \frac{1}{\Delta_{b\beta}} \,
      {\cal Z}_{\beta} )^{st}.
     \end{equation}
In order to express $W_{b\beta}$ in terms of $W_{a\alpha}$ we first
note that the derivatives become
     \begin{eqnarray}
		 D_{b} & = & e^{X}D_{a}    \label{G6}\\
		 \partial_{b} & = & e^{2X}(\partial_{a}+(D_{a}X)D_{a})
	\label{G7}
     \end{eqnarray}
where we have set $e^{-X}=D_{a}\theta_{b}$ (which is the canonical 1-cocycle
of $S\Sigma$ \cite{CR}).
Then it is straightforward to build the transformation matrix $T_{ba}$
defined by:
\[   {\cal D}_{b}=T_{ba} \cdot {\cal D}_{a}.
\]
The determinant $\Delta$ of $W_{0}$ transforms as
\[   \Delta_{b \beta}=e^{nX}\frac{\Delta_{a \alpha}}{(\Phi^{0}_{\beta\alpha},
{\cal Z}_{\alpha})_{E}}.
\]
From the definition (\ref{tr}) it follows readily that
$$
\Phi_{\alpha\alpha}=\unit ,
$$
$$
\Phi_{\alpha\beta}\Phi_{\beta\gamma}\Phi_{\gamma\alpha}=\unit .
$$
Thus $\Phi$ defines a flat \sln\ vector bundle on $S\Sigma$
\footnote{In general one has $\Phi_{\alpha\alpha}=c_{\alpha}\unit$ and
$\Phi_{\alpha\beta}\Phi_{\beta\gamma}\Phi_{\gamma\alpha}=
k_{\alpha\beta\gamma}\unit$, where $c_{\alpha}$ and $k_{\alpha\beta\gamma}$
are constants. However since sdet($\Phi$)=1 the result above follows.}.
We then have : $W_{b \beta}=T_{ba}K_{ba}W_{a \alpha}
\Phi^{st}_{\beta \alpha}$
where $K_{ba}$ has the form
\[   \left( \begin{array}{cc}
		  K_{0} & 0 \\
		  K_{1} & K_{2}
	    \end{array}
     \right)
\]
with
\begin{itemize}
\item for $0\leq i\leq n$ and $j\leq i$
	$$
(K_{0})_{ij}=\left( \begin{array}{c} i \\ j \end{array} \right)
\partial^{i-j} e^{-nX}
$$
\item for $n+1\leq i\leq 2n$ and $j\leq i$
$$(K_{1})_{ij}= \left( \begin{array}{c} i-n-1 \\ j \end{array} \right)
D\partial^{i-n-j-1} e^{-nX}
$$
\item for $n+1\leq i\leq 2n$ and $n+1\leq j\leq i$
$$
(K_{2})_{ij}=\left( \begin{array}{c} i-n-1 \\ j-n-1 \end{array} \right)
\partial^{i-j} e^{-nX}.
$$
\end{itemize}
From now on we set : $\Lambda_{ba}=T_{ba}K_{ba}$.
This matrix can be viewed as
a transition function of a bundle over $S\Sigma$, namely the jet bundle. We
recall its definition. A $n$-jet of a field $\psi$ of weight $\f{n}$
(i.e. of a section of the canonical bundle over $S\Sigma$) is the
vector field
$$ j_{n}\psi=(\psi,\partial\psi,...,\partial^{n}\psi,D\psi,...,
D\partial^{n-1}\psi)^{st}.$$
Under a superconformal change of coordinates we have
$$ j_{n}\psi_{b}=\Lambda_{ba}j_{n}\psi_{a}.$$
Thus writing
 $$\Phi^{\vee}_{ \alpha \beta }=W_{a \alpha}^{-1}
		      \Lambda_{ab} W_{b \beta} $$
where
$$\Phi^{\vee}_{ \alpha \beta }=\Phi^{st\ -1}_{\alpha \beta} $$ is the dual
bundle,
amounts to saying that the jet bundle $\Lambda$ and the flat \sln\ bundle
$\Phi$ are equivalent.
We can now define the two matrices
     \begin{equation}
	    \om=DW \cdot W^{-1}  \label{om}
     \end{equation}
     \begin{equation}
	    \os=\bar{D}W \cdot W^{-1} . \label{omstar}
     \end{equation}
The crucial property of these matrices is their independence on
the choice of the index $\alpha$, i.e. on the choice of a chart in the
super-projective atlas. These super-matrices are odd and transform as
\begin{eqnarray}
 \om_{b} & = & e^{X}[\tilde{\Lambda}_{ba}\om_{a}\Lambda^{-1}_{ba}+
(D_{a}\Lambda_{ba})\Lambda^{-1}_{ba}] \label{12} \\
 \os_{b} & = & e^{\bar{X}}\tilde{\Lambda}_{ba}\os_{a}\Lambda^{-1}_{ba}
\label{13}
\end{eqnarray}
where the tilde means that the odd blocks of the matrix acquire a minus sign.
It is straightforward to verify that $\om$ and $\os$ satisfy the
following relation
 \begin{equation}
     \bar{D}\om+D\os+\tilde{\om}\os+\tilde{\om}^{\ast}\om=0 .
	   \label{holo}
 \end{equation}
Furthermore we have
 \begin{eqnarray}
     str\om & = & 0 \label{trace} \\
     str\os & = & 0   .
 \end{eqnarray}
The three last equations indicate that $\om$ and $\os$ can be viewed
as the two components of a flat \sln\ connection on the jet bundle
$\Lambda$.

Not all the coefficients of $\om$ and $\os$ are independent. In
fact there are only $2n$ independent fields for each matrix.
The peculiar structure itself of the matrix $W$ (i.e. a Wronskian structure)
 entails the relations
\begin{equation}
DW_{ij}=\left\{ \begin{array}{ll}
		    W_{i+n+1,k}  & \mbox{for $0 \leq i \leq n-1$} \\
		    W_{i-n,k}  & \mbox{for  $n+1 \leq i \leq 2n$}
		   \end{array}
	   \right.     \label{deriv}
\end{equation}
and therefore
\[ \om_{ij}=\left\{ \begin{array}{ll}
			\delta_{i+n+1,j} & \mbox{for $0 \leq i \leq n-1$} \\
			\delta_{i-n,j} & \mbox{for  $n+1 \leq i \leq 2n$}
		       \end{array}
	      \right.
\]
The only remaining coefficients are the $\om_{n,j}$. But due to the
relation (\ref{trace}) we also have $\om_{n,n}=0$ leaving us with $2n$
independent fields $\om_{n,j} \;\; j \neq n$. Thus the matrix $\om$,
in this block grading, looks like
\begin{equation}
\om=\left( \begin{array}{ccccccc}
	0 & \cdots & \cdots  & 0 & 1 & &  \\
	\vdots &   &  & \vdots	&  & \ddots &	\\
	0 & \cdots & \cdots  & 0  &  &	& 1 \\
	\om_{n,0} & \cdots & \om_{n,n-1} & 0 &
			    \om_{n,n+1} & \cdots & \om_{n,2n} \\
	0 & 1 & \cdots & 0 & 0 & \cdots & 0 \\
	\vdots	&  & \ddots &  & \vdots & & \vdots  \\
	0 & \cdots & 0 & 1 & 0 & \cdots & 0
	\end{array}
	\right)
\label{ombloc}
\end{equation}
As regards $\oms$ it is easier to use a grading by diagonals
\cite{theisen,gt} which is moreover better suited for the following
applications, in particular for dealing with the root vectors
of the Lie algebras.
Of course there are several ways to transform the matrix from the block
to the diagonal
grading. Here we impose that all the 1 entries in $\om$ be gathered
on the first
diagonal below the main one.
The matrix which permits one to pass from block to diagonal grading
is defined
by
$$
	M_{diag}=P^{-1}M_{block}P\ \ \ \ \ \
	\mbox{and}\ \ \ \ (P)_{ij}=\delta_{ip(j)}
$$
where $p$ is the permutation given by :
 $$
\begin{array}{cc}
		   p(2k+1)=2n-k & \ \ \ \ \ p(2k)=n-k. \\
\end{array}
$$
Thanks to this grading we can decompose \sln\ in a sum of subspaces $B_{i}$,
in which
each subspace corresponds to a diagonal. We choose to number these
subspaces with respect to the main diagonal which will be the $0^{th}$ one.
Positive numbered diagonals will be located on the upper triangular part of
the matrices and negative ones on the lower part:
 $$sl(n+1\!\mid\!n)=\bigoplus_{s=-2n}^{2n} B_{s}
		   =B_{-}\oplus B_{0} \oplus B_{+}.$$
In this new grading the matrix $\om$ has the form
\begin{equation}
 \om= \left(
      \begin{array}{ccccc}
 0 & \rho_{2} & \rho_{3} & \cdots &  \rho_{2n+1} \\
 1 & 0 & & &  \\
   & \ddots & \ddots & & \\
   &  & \ddots & \ddots &  \\
      &  & & 1 & 0
	\end{array}
	\right)
\end{equation}
where the $\ro{i}$ field is the $\om_{n,k}$ of conformal weight
$\theta^i(\ro{i}\equiv \ro\mid_{\theta^i}$). We note that
this form of $\om$ is reminiscent of the Polyakov's partial-gauge fixed
connection \cite{Pol2}.

The flatness condition (\ref{holo}) contains the $2n$ super-holomorphy
conditions obeyed by the $\ro{i}$. These conditions are in fact the Ward
identities of the induced super $W$-gravity model underlying this
geometrical framework \footnote{The association of a zero curvature
condition to the formulation of induced $W_n$ gravity and its interpretation
as an anomaly equation are not new and can be found in numerous earlier
works (see for instance \cite{bfk},\cite{Das}).}.
Furthermore, when $\Db \om = 0$, it allows us
to determine the elements of $\oms$ in terms of these $\ro{i}$ and of $2n$
other independent superfields $\mmu{i+1} \equiv
\mu\mid_{\tbar}^{\,\theta^{i}} \equiv\oms_{i,0},\  i\neq 0$.
This set of relations can be treated in three groups corresponding to the
components of $\oms$ in $B_-$, $B_0$ and $B_+$. For the first group we
solve iteratively the equations  for $s=j-i$ going from
 $-2n$ to $ -2$, with, for each value of $s$, $j$ running from 0 to $2n+s$.
This provides us with all the entries below the main diagonal. The second
group concerns the elements  of the $0^{th}$ diagonal : they are obtained
for $s=-1$ and $j$ going from $0$ to $2n-1$, with the help of the condition
(\ref{trace}).
At last  $0 \leq s \leq 2n-1$ with $j$ running from $2n$ to $s$ gives the
elements above the main diagonal\footnote{We are there in the situation
referred to in \cite{BG} where $\om$ is composed of a constant part (which
will be named $J_-$ in the following section) and a part which contains the
fields $\ro{i}$; moreover the algebra is graded and $J_-$ has a definite
degree, namely $-1$. This is why elementary
calculations lead to the results and it is not necessary to use the more
elaborated and more general method of \cite{BG}.}. The results are
\[
\begin{tabular}{lcrcl}
 for $s=-2n,\cdots, -2, $& & & & \\[5mm]
\multicolumn{2}{c}{$ j = 0 $}&$ \oms_{-s,1} $&$ = $&$ (-1)^{-s+1}
(\mmu{-s} + D\mmu{-s+1})$ \\[3mm]
  \multicolumn{2}{c}{$ j = 1,\cdots,2n+s  $}&$ \oms_{j-s,j+1}  $
&$ = $&$ (-1)^{1-s} (D\oms_{j-s,j}+\oms_{j-s-1,j})
+ (-1)^{j}\mmu{j-s+1}\ro{j+1}$,  \\[5mm]
for $ s=-1,  $\\[5mm]
&  &$ \oms_{0,0} $&$=$&$ \sum_{p=1}^{n}(\mmu{2p+1}\ro{2p}
- D\oms_{2p,2p-1})$ \\[3mm]
  & &$ \oms_{j,j} $&$ = $&$ \sum_{p=2}^{j}(D\oms_{p,p-1}
  +(-1)^{p+1} \mmu{p+1}\ro{p})+ \oms_{0,0}
+ D\mmu{2}, $ \\[5mm]
for $s = 0,\cdots,2n-1,  $& & & & \\[5mm]
\multicolumn{2}{c}{ $  j = 2n  $}&$ \oms_{2n-1-s,2n} $&$ = $&$
(-1)^{s} \mmu{2n+1-s}\ro{2n+1} - D\oms_{2n-s,2n}$ \\[3mm]
 \multicolumn{2}{c}{$ \hspace*{1mm}  j = 2n-1,\cdots,s+1  $}&$
\oms_{j-s-1,j} $&$ = $&$
  (-1)^{1-s}\oms_{j-s,j+1}+(-1)^{j-s}\mmu{j-s+1}\ro{j+1} - D\oms_{j-s,j}$
\end{tabular}
\]\\[5mm]
$\oms$ in this diagonal grading looks as follows
\begin{equation}
\label{matdiag}
 \oms= \left(
      \begin{array}{ccccc}
 *& * & \cdots & * &* \\
 \mmu{2} & * & \cdots & * & * \\
 \mmu{3} & * & \cdots & * & * \\
 \vdots & \vdots & \cdots & \vdots & \vdots \\
 \mmu{2n+1} & * & \cdots & * & *
	\end{array}
	\right),
\end{equation}
where the stars stand for expressions in terms of the $\mmu{i}$.

Thus to the family of generalized superprojective structures
$\{(Z^1,\ldots,Z^n;\Theta^1,\ldots,\Theta^n)\}$ on $S\Sigma$
is canonically
associated a set of $2n$ pairs of geometrical fields $(\mmu{i},\ro{i})$
which can be viewed as the generalized super-Beltrami
differentials and
as the generalized projective connections (i.e. the backgrounds fields),
respectively,
in the same way that usual projective structures
are parametrized by the Beltrami coefficient and the Schwarzian derivative.
These sets of fields contain the supersymmetric extension
$H^{\ z}_{\bar{\theta}}$ of
the ordinary Beltrami coefficient \cite{dgbel} and the superprojective
connection.
In fact this parametrization of the generalized projective structures
in terms of the $(\mmu{i},\ro{i})$ is one-to-one. Indeed
starting from the pairs $(\mu_i,\rho_i)$ and
defining the matrices above we can obtain generalized
superprojective structures canonically associated to the $\mu_i$'s and the
$\rho_i$'s. The equation (\ref{holo}) can be viewed as an
integrability condition for a linear system of partial differential
equations
\footnote{Thus this geometrical construction provides us with a Lax pair
for the super Ward identities as integrability conditions.}
\begin{equation}
(D-\om)U=0  \hspace{2cm}    \hspace{2cm}   (\bar{D}-\os)U=0,
\label{axlp}
\end{equation}
with the constraint
\begin{equation}
  sdetU=1.
\label{detu}
\end{equation}
To parametrize the matrix $U$ we consider a set of local sections
(differentials) transforming homogeneously
\[
 \Psi = \left(\begin{array}{ccccccc}
\Psi_0,  &  \cdots, &  \Psi_{2n}
\end{array}
\right)
\]
and define
$$U_{ij}=D^{i}\Psi_j \ \ \ \ \ 0\leq i\leq 2n,$$
Then the first equation (\ref{axlp}) is
equivalent to a set of differential equations
 ${\cal L}\Psi_j=0$, $0\leq j\leq 2n$, where
\begin{equation}
  {\cal L}=D^{2n+1}-\sum_{l=1}^{2n}(\om_{n,l}D^{2n-l})	    \label{lax}
\end{equation}
is the Lax operator in the ($n$-reduced) super-KP hierarchy. Such a system
admits $2n+1$ linearly independent local solutions $(\Psi_0,...,\Psi_{2n})$
which are normalized so that (\ref{detu}) holds.
The non-uniqueness is reduced by imposing that the $\Psi_i$'s satisfy the
second equation (\ref{axlp}), and the zero-curvature condition (\ref{holo})
is now regarded as the compatibility equation of the linear system
(\ref{axlp}). For instance let us assume that $\Psi_{2n}$ is nowhere
vanishing so that the maps
$Z^{n-i}=\frac{\Psi_{2i}}{\Psi_{2n}}$ and $\Theta^{n-i}=\frac{\Psi_{2i+1}}
{\Psi_{2n}}$ with
$1\leq i\leq n$ are well defined; it is then easy to verify that these maps
satisfy eqns. (\ref{Moebius},\ref{tr}).

\section{Classical super $W$-algebras}

To obtain the $W$-algebras we consider the independent fields appearing in
the connection defined in the preceding section. These fields are in general
not
covariant under a super-conformal change of the coordinates $\sz=(z,\theta)$.
The first step is to make a basis change for $\om$ in order to find new
fields that transform homogeneously \cite{ds}. For this purpose we shall
extend to the supersymmetric case the method of \cite{zuber} (see also
\cite{bersh} and \cite{sevrin}).
Then thanks to the connection
 we define a symplectic form which allows us to covariantize the fields
appearing in $\os$.
\subsection{The Drinfeld and Sokolov method in the supersymmetric case}

As noted in the preceding section, due to its form the matrix $\om$ can
be thought of as
the matrix appearing in a first order matrix differential equation. If we
write $\hat{\cal L}F=(D-\om)F=0$
with $F=(f_{0},f_{1},\ldots,f_{2n})^{st}$, we can eliminate the $f_{i}$'s to
obtain a differential operator of the form \cite{gt}
\begin{equation}
 {\cal L}=D^{2n+1}-\sum_{k=1}^{2n}\rho_{k+1}D^{2n-k} \label{op}
\end{equation}
where $\rho_{k}=\om_{0,k-1}$ .
We mention here that this operator can be factorized as follows
$$ {\cal L}=(D-D\Phi_{2n})(D-D(\Phi_{2n-1}-\Phi_{2n}))...
			(D-D(\Phi_{1}-\Phi_{2}))(D-D\Phi_{1}) $$
leading to new fields $\Phi_{i}$ that obey Toda equations under the zero
curvature condition (\ref{holo}). This field redefinition is a generalized
Miura transformation \cite{noh}.

$\om$ can be decomposed as $J_{-}+R $ where $J_{-}=diag_{-1}(1,...,1)$.
$J_{-}$ can be identified with the sum of the negative root vectors $f_{i}$
of \sln\ (see appendix). It can also be regarded as one of the generators
of an \osp\ algebra \cite{frs}. Indeed defining
$$ J_{+}=\sum_{i=1}^{r}\sum_{j=1}^{r}(C^{-1})_{ij}e_{i} $$
where $C$ and $r=2n$ are the Cartan matrix and the rank of \sln\, respectively,
and
$$ H=\{ J_{+},J_{-} \} \ \ \ \ \ X_{\pm}=\f{1}\{ J_{\pm},J_{\pm} \}$$
we have the following commutation relations
\begin{equation}
 \begin{array}{ll}
   [ H,J_{\pm} ]=\pm J_{\pm}   &   [ H,X_{\pm} ]=\pm 2 X_{\pm}	 \\

  [ X_{\pm}, J_{\mp} ]= \mp J_{\pm}   &   [ X_{\pm}, J_{\pm} ]=0
  \end{array}
\label{osppal}
\end{equation}
$$  [ X_{\pm},X_{\mp} ]=-H. $$
The matrices $H$, $J_{\pm}$ and $X_{\pm}$ generate an \osp\ subalgebra
\cite{nahm}
and $H$ allows us to characterize the diagonal grading since, as can be
straightforwardly verified, for any $M$ in $B_{i}$, we have \cite{theisen,gt}
\begin{equation}
 [H,M]=iM  .\label{chardiag}
\end{equation}
Going back to the operator $\cal{L}$ it is known
\cite{ds,zuber} that $\hat{\cal L}$ is not the only matrix
operator leading to $\cal{L}$ (\ref{op}). In fact every
 $\hat{\cal L}'=\tilde{Q}\hat{\cal L}Q^{-1}$
(where $Q$ is a coordinate dependent upper triangular matrix with 1 entries
on the main diagonal) will lead to the same ${\cal L}$.
Writing $ Q=\unit+\sum_{i \geq 1} Q_{i} $  , $Q_i \in B_i$ we easily see that
we can obtain
$$ \hat{\cal L}'=D-J_{-}-R' $$
where $R' \in B_+$ (like ($\om - J_-$)) provided $\{Q_1,J_-\}=0$, which leads
to $Q_1 = 0$.
This gauge freedom on $\hat{\cal L}$ is used to
reparametrize $\cal L$ in terms of new fields $\tilde{\rho_{i}}$ which are
covariant under a super-conformal change of coordinates.
To get this new $\hat{\cal L}'$
following \cite{ds} we decompose every $B_{i}$ as :
$$ B_{i}=\left[ J_{-},B_{i+1} \right] \oplus V_{i} $$
where [.,.] is the graded commutator. The map $[J_{-},\cdot ]$ from $B_{i+1}$
to $B_{i}$ is injective (for \sln\ ), so that
$$ dim\left[ J_{-},B_{i+1} \right]=dimB_{i}-1 $$
and  $$ dimV_{i}=1. $$
We now require the matrix $R'$ to belong to $V=\oplus V_{i}$.
There remains the task of choosing a convenient
basis for the $V_{i}$'s. To do this we select a matrix
$J_{1}$ in $V_{1}$ and use its powers $J_{1}^{k}$ as a basis for $V_{k}$
(We note that this choice corresponds to the highest weight gauge
of \cite{wipf} ). Unlike the bosonic case where the choice is
canonical \cite{zuber} we cannot choose $J_{+}$ as a basis in $V_{1}$ since
 $J_{+}=\left[ J_{-},X_{+} \right]\notin V$. We thus have to impose
constraints on $J_{1}$ so that the matrix operator $\hat{\cal L}'$ be
 written
\begin{equation}
 \hat{\cal L}'=D-J_{-}-\sum_{k=1}^{2n}\rt{k+1}J_{1}^{k} \label{opL}
\end{equation}
where the new coefficients $\rt{k}$ are covariant of weight
$\f{k}$
      \footnote{A field of weight $\f{h}$ transforms as
      $\psi(z',\theta')(D\theta')^{h}=\psi(z,\theta)$ i.e. in infinitesimal
 form $\delta_{\epsilon}\psi=\left(\epsilon\partial+\f{1}(D\epsilon)D+\f{h}
			(\partial\epsilon)\right)\psi $ }
except $\rt{3}$ which transforms as a projective connection, i.e. under an
infinitesimal change of coordinates we have
\begin{equation}
 \delta_{\epsilon} \hat{\cal L}' =\left[ \chi, \hat{\cal L}' \right] =
	\f{1}(D^{5}\epsilon)J_{1}^{2}+
	\sum_{k=1}^{2n} \left(\epsilon\partial+\f{1}(D\epsilon)D+\f{k+1}
	(\partial\epsilon)\right)\rt{k+1}J_{1}^{k}  \label{cooch}
\end{equation}
since under an infinitesimal superconformal change of coordinates
$$ \left\{ \begin{array}{lll}
		z'& = & z+\epsilon+\f{1}(D\epsilon)\theta \\
		\theta' & = & \theta + \f{1} (D\epsilon)
	 \end{array}
  \right.
$$
the super Schwarzian derivative
$S=\frac{\partial^{2}\theta'}{D\theta'}
      -2\frac{(\partial\theta')(\partial D\theta')}{(D\theta')^{2}} $
transforms as
$$ \delta_{\epsilon} S=\left( \epsilon\partial+\f{1}(D\epsilon)D+\f{3}
	(\partial\epsilon) \right)S+\f{1}(D^{5}\epsilon). $$
The coordinate dependent matrix $\chi$ generates the transformation and
is such that when applied
on a vector $F=(f_{0},f_{1},\ldots,f_{2n})^{st}$ the lowest component
$f_{2n}$ transforms as a covariant field of weight $-\f{n}$
(since $\cal{L}$ is covariant
when applied on fields of weight $-\f{n}$ \cite{zuber,gierop}).
Expanding $\chi$ on the \osp\ basis we obtain the following constraints
$$ [J_{+},J_{1}^{k}]=0, \ \ \ [J_{1}^{2},J_{-}]=-J_{+}. $$
The unique solution to these constraints (up to a sign) is
$$ J_{1}=diag_{+1}(n,1,n-1,2,...,n) $$
which is an operator obtained in \cite{gt}
and the diffeomorphism generating operator $\chi$ is found to be
$$-\chi=\f{1}\left( (D\epsilon)(J_{-}+R)
	+(\partial\epsilon)H+(D^{3}\epsilon)J_{+}
	+(D^{4}\epsilon)X_{+}+2\epsilon(J_{-}+\tilde{R})(J_{-}+R)
	+2\epsilon(DR) \right) . $$

\subsection{Covariantization of the fields}
The Drinfeld and Sokolov method thus leads to covariant fields $\rt{k}, \
k \neq 3$.
To obtain explicitly these $\rt{k}$ and the matrix $Q$ we just have to
compare $\tilde{Q}\hat{\cal L}$ with $\hat{\cal L}'Q$ in every subspace
$B_{i}$. For the first ones we have
\begin{eqnarray*}
 \rho_{2} & = & n(n+1)\rt{2} \\
 \rho_{3} & = &\f{1}n(n+1)(\rt{3}+D\rt{2}).
\end{eqnarray*}
Since $Q$ is expressed in terms of the fields $\rho_{k}$ and their
derivatives the gauged fields $\rt{k}$ contain non-linear combinations of
these fields.
Under a change of coordinates the field $\rt{3}$ transforms as
a projective connection i.e.
\begin{equation}
 \rt{3}'=e^{3X}(\rt{3}+S) \label{Tschw}
\end{equation}
where $S$ is the super-Schwarzian derivative.

A change of basis for $\hat{\cal L}$ generates the following transformations
\begin{eqnarray}
\om' & = & \tilde{Q}\om Q^{-1}+DQ\cdot Q^{-1}  \label{omq} \\
\Omega^{\ast'} & = &
\tilde{Q}\os Q^{-1}+\bar{D}Q\cdot Q^{-1}. \label{omr}
\end{eqnarray}
To covariantize the $ \mmu{k}$ fields we introduce the symplectic form
\cite{fk}
\begin{equation}
\omega=\int_{S\Sigma}str(\delta\os\wedge\delta\omt) \label{sympl}
\end{equation}
defined on the manifold $S{\cal M}$ of connection one-forms $\con$
\begin{equation}
\con=\om d\sz+\os d\szb,
\label{scon}
\end{equation}
where $d\sz=(dz\!\mid\! d\theta$),
the operator $\delta$ in (\ref{sympl}) being the exterior derivative on
$S{\cal M}$. The equation (\ref{holo}) is nothing but the flatness
condition ${\cal F}=0$ for $\con$.
This symplectic 2-form is well defined on $S\Sigma$ since
the operator $\delta$ is inert with respect to a coordinate change
$\sz_{a}\longrightarrow \sz_{b}$; namely the gluing properties
(\ref{12}) become
\begin{eqnarray}
\delta\om_{b} & = &
	 e^{X}\tilde{\Lambda}\delta\om_{a} \Lambda^{-1} \label{omqd} \\
\delta\os_{b} & = &
 e^{\bar{X}}\tilde{\Lambda}\delta\os_{a} \Lambda^{-1}. \label{omrd}
\end{eqnarray}
Using the explicit expression of $\om$ and $\os$ (\ref{matdiag}) we can write
$$\omega=\int_{S\Sigma}\sum_{k=1}^{2n} \delta\mu_{k+1}
				 \wedge\delta\rho_{k+1}, $$
this replacement being equivalent to the Hamiltonian reduction of \cite{bfk}
and \cite{boo} so that $S{\cal M}$ is in fact the reduced manifold.
Letting $\delta$ act on $\ro{k+1}$ expressed in terms of the $\rt{j}$ leads to
a polynomial which is linear in $\delta\rt{j}$ and possibly in its derivatives
$\delta D^l\rt{j}$. Then, with the help of integrations by parts, we can
factorize $\delta\rt{k+1}$, getting in this way the expression of
$\delta\mt{k+1}$ :
$$\omega=\int_{S\Sigma}\sum_{k=1}^{2n} \delta\mu_{k+1}
				 \wedge\delta\rho_{k+1}
     =\int_{S\Sigma}\sum_{k=1}^{2n} \delta\mt{k+1}
				 \wedge\delta\rt{k+1} .$$
Since the covariantized field $\mt{k+1}$ does not depend explicitly on the
superprojective coordinates, we obtain straightforwardly $\mt{k+1}$ from
$\delta \mt{k+1}$
Thanks to the fact that the integrand in $\omega$ is well-defined on a SRS
the $\mt{k}$ transform as
$$ \mt{k}'=e^{\bar{X}}e^{(1-k)X}\mt{k}. $$
An explicit example of this construction is given in sect. 5.1.
Note that although $\rt{3}$ transforms as a projective connection $\mt{3}$
is covariant. Indeed when one applies the functional
exterior derivative $\delta$ to (\ref{Tschw}) the term $S$ does not contribute
since $\delta S=0$.

Let us now study  the algebra formed by the $\rt{i}$ and
analyze its spin content. Considering the operator
$\hat{\cal{L}}$ we immediately see that we have spins
$$ (1,\f{3},2,\f{5},...,n,n+\f{1})$$
or equivalently considering the physical component
expansion \footnote{we have : $\rt{i}=\rt{i}_{0}+\theta \rt{i}_{1}$}
$(\rt{i}_{0},\rt{i}_{1})$
$$ \left((1,\f{3});(\f{3},2);(2,\f{5});...;(n,n+\f{1});(n+\f{1},n+1)\right).$$
Obviously we are dealing with the $N=2$ super $W_{n+1}$ algebra
since the spins naturally gather in $N=2$ supermultiplets this fact
being intimately related to the \sln\ algebra. Now as it is well known
\cite{theisen,drs} we can restrict our connection form to belong to a
subalgebra of \sln\ namely $osp(2m\pm1\!\mid\!2m)$
(with $4m\pm1=2n+1$) and find a Chevalley basis of these subalgebras which
gives the same expressions
for the generators of the \osp\ subalgebra (\ref{osppal}) leading to a very
convenient characterization of the basis vectors $J_{1}^{k}$ that belong to
$osp(2m\pm1\!\mid\!2m)$. Indeed they are such that
$k=2,3\ \mbox{mod}\ \ 4$. The spin content is now restricted to
$$ \left((\f{3},2);(2,\f{5});(\f{7},4);(4,\f{9});
		...;(2n-\f{1},2n);(2n,2n+\f{1})\right).$$

\section{The BRST symmetry and the consistent anomaly}

The main advantage of this geometrical framework is to define the
$SW_n$ symmetries as gauge transformations of the vector bundle $\Phi$
and to
provide a systematic method to derive a nilpotent BRST
algebra, as we now discuss. This is a straightforward extension of the
framework proposed by Zucchini \cite{zucchini} to formulate the symmetries
of the induced light cone $W_n$-gravity. The most used path to study the
quantum invariance of these theories consists in deriving from the underlying
algebra the BRST charge Q. The knowledge of this operator is essential in a
great body of work in W strings \cite{Pope} towards unravelling the spectrum
of physical states. The failure of Q$^2$ to vanish leads to an anomaly in the
BRST operator algebra. However the obtention of anomalies in the BRST
Ward identities requires the construction of a nilpotent BRST algebra. The
two approaches are difficult to compare although in a recent paper
\cite{MPSX} a relation between these two notions has been noticed.

\subsection{BRST algebra}

In sect.2 we have seen that the transition functions on overlapping domains
of the local maps ${\cal Z}^i_{\alpha} \equiv (Z^i_{\alpha},\Theta^i_{\alpha})$
define an \sln\ -valued 1-cocycle $\Phi_{\alpha \beta}$ on $S \Sigma$ which in
 turn corresponds to a flat \sln\ vector bundle $\Phi$ on $S \Sigma$. Such
bundle is
canonically associated to a generalized projective structure and can be
considered as a functional of the fields ($\rt{i},\mt{i}$). The variations of
these fields which leave this bundle invariant are precisely the form of the
super $W_n$-symmetry transformations. These variations are obtained from
deformations of the maps ${\cal Z}^i$ which are defined by

\begin{equation}
{\cal Z}'=\frac{R\cdot {\cal Z}}{(R^{2n},{\cal Z})_{E}},
\label{va1}
\end{equation}
where $R$ is an OSp($n+1 \mid\!\!n$) matrix, and $R^{2n}$ is the
$(2n+1)^{th}$ row (labelled $2n$) of $R$
($R$ being written in diagonal grading).
Requiring for consistency, that ($Z,\Th$) and ($Z',\Th '$) glue on
overlapping domains as in (\ref{tr}) i.e.

\begin{equation}
 R_{\alpha} \Phi_{\alpha \beta} = \Phi_{\alpha \beta} R_{\beta},
\label{va2}
\end{equation}
means that these coordinates are related by a gauge transformation of
the flat vector bundle $\Phi$  defined by the matrix function $R$.
The infinitesimal variations of the maps given in terms of infinitesimal
parameters $\epsilon^i_j$ are
\begin{equation}
\delta {\cal Z}^i =\epsilon^i_k {\cal Z}^k-\epsilon^{2n}_k {\cal Z}^k
{\cal Z}^i .
\label{va3}
\end{equation}
These transformations are generalization of the laws given by the group
OSp($2 \mid\!\!1$) and which are the infinitesimal form of the well-known
 superconformal transformations \cite{Nin} for the $N=1$ SRS, namely

\begin{eqnarray}
Z^{\prime}&=&\frac{aZ+b}{cZ+d}+ \Th \frac{\alpha Z+\beta}{(cZ+d)^2},
   \label{11a}\\
\Th^{\prime}&=&\frac{\alpha Z+\beta}{cZ+d}
+ \Th\frac{1}{cZ+d}.
  \label{12a}
\end{eqnarray}
with $R$ belonging to OSp(2$\mid\,$1), i.e.
$$R=\left(
\begin{array}{ccc}
a &  \alpha b-\beta a & b\\
\alpha & 1-\alpha\beta&\beta\\
c & \alpha d -\beta c & d
\end{array}
\right), \ \ ad-bc=1+\alpha\beta .
$$
Indeed, in infinitesimal form these transformations read

\begin{eqnarray}
\delta Z &=& \epa +2 \epo Z -\epb Z^{2} -\epe \Th - \epd \Th Z,
  \label{13a}\\
\delta \Th &=&\epd Z + \epe +\epo \Th  -\epb \Th Z,
   \label{14a}
\end{eqnarray}
where the infinitesimal parameters $\epi$, $i=0,1,2$ and the $\epd,\epe$
are Grassmann even and odd respectively. For the sake of comparison let us
restrict ourselves to the Wess-Zumino gauge (W-Z) where $DZ= \Th D \Th$. If
we introduce the infinitesimal parameter
\begin{equation}
\Upsilon =  (\epa +2 \epo Z -\epb Z^{2} - 2 \epe \Th - 2\epd \Th Z)
\frac{1}{(D \Th)^2},
\label{15a}
\end{equation}
and assume for simplicity that the $\epsilon_i$'s are antiholomorphic,
these variations become

\begin{eqnarray}
\delta Z &=& \E(D \Th)^2-\Th \delta \Th,
\label{sb1}\\
\delta \Th &=&	\f{1} D \E D \Th +\E \drp \Th .
\label{ss2}
\end{eqnarray}
 They are identical to the BRST laws given in \cite{dgbel} which read
\begin{eqnarray}
sZ  & = & C^z(D \Th)^2 - \Th s \Th, \label{ss3} \\
s\Th & = & \f{1} D C^z D \Th +C^z \drp \Th,  \label{ss4}
\end{eqnarray}
when the infinitesimal parameter $\E$ has been turned into the ghost field
$C^z$ and the gauge transformation $\delta$ into the BRST operator $s$.
These laws give the forms of the BRST transformations
of the superprojective connection and of the super-Beltrami differential,
which are the
current and the gauge superfield respectively of the induced supergravity in
the W-Z gauge.
Thus the well-known results of the $N=1$ SRS \cite{dgbel}
restricted to this gauge are easily reproduced.

Now from the construction of the generalized superprojective connections and
super-Beltrami differentials given in sect.2, and thanks to the
transformations (\ref{va3}) we generalize the above result to the formulation
 of the nilpotent BRST algebra corresponding to the classical $SW_n$
symmetry.

The super determinant $\Delta '$ of $W_0'$ obtained by replacing in
(\ref{w0}) the ${\cal Z}$'s by the ${\cal Z}'$'s defined by eq.(\ref{va1})
 is
\begin{equation}
\Delta '=\frac{\Delta}{(R^{2n},{\cal Z})_{E}} sdet\Bigl([\kappa^{(l)}
\tilde{W_0}D^{(2l+1)} (R^{st}) +\kappa^{(l) \prime}
W_0 D^{(2l)}(R^{st})]W^{-1}_0\Bigr).
\label{ddt}
\end{equation}
where $\kappa^{(l)}$ and $\kappa^{(l) \prime}$ denote numerical matrices
 which are respectively defined by

$$\kappa^{(l)}_{2q+1,p}=C^l_{n-1-q} \delta_{p,2(q+l+1)} ; \hspace{.2in}
\kappa^{(l)}_{2q,p}=0.$$

$$\kappa^{(l) \prime}_{2q+1,p} =C^l_{n-1-q} \delta_{p,2(q+l)+1} ;
\hspace{.2in}
\kappa^{(l) \prime}_{2q,p} =C^l_{n+q} \delta_{p,2(q+l)}.$$
It then follows that

\begin{equation}
[\kappa^{(l)}, \kappa^{(m)}] =[\kappa^{(l)}, \kappa^{(m) \prime}]
= [\kappa^{(l) \prime}, \kappa^{(m) \prime}] =0.
\label{cm}
\end{equation}
From the definition of $W$ in terms of $W_0$ (given by eqs.(\ref{w0},
\ref{defw})) it is easy to show that

\begin {equation}
W=\Xi W_0,
\label{psi}
\end{equation}
where the matrix $\Xi$ can be written in terms of $\kappa$ and
$\kappa^{\prime}$
\beqna
\Xi_{q,2l} & = &
\sum_{p=0}^{2n}\kappa^{(n-l) \prime}_{q,2n-p} D^{(p)}(\Delta^{-1}),
 \nonumber  \\
\Xi_{q,2l+1} & = & \sum_{p=0}^{2n}\kappa^{(n-1-l)}_{q,2n-p}D^{(p)}(\Delta^{-1}).
\nonumber
\eeqna
We can verify that
\beqna
\Xi\kappa^{(l) \prime}= \kappa^{(l) \prime }{\Xi}
\quad \mbox{and} \quad
\Xi{\kappa^{(l)}}= \kappa^{(l)}\tilde{\Xi}
..
\eeqna
By using these relations, one can replace $W_0$ and $\tilde{W_0}$ by $W$ and
$\tilde{W}$ respectively in (\ref{ddt})
\begin{equation}
\Delta '=\frac{\Delta}{(R^{2n},{\cal Z})_{E}} sdet\Bigl([\kappa^{(l)}
\tilde{W}D^{(2l+1)} (R^{st}) +\kappa^{(l) \prime}  W D^{(2l)}(R^{st})]
W^{-1}\Bigr).
\label{ldt}
\end{equation}
In order to obtain the BRST algebra corresponding to (\ref{ldt}), let us
consider the infinitesimal parametrization of these transformations by
setting $R=1+ \epsilon$ (with str($\epsilon$)=0). This linearizes the r.h.s
of (\ref{ldt}) which becomes

\begin{equation}
\delta \Delta =-\Delta \biggl[(\epsilon^{2n},{\cal Z})_{E}-
str\Bigl(\bigl(\sum_{l=0}^{n-1}\kappa^{(l)} \tilde{W} D^{(2l+1)}
(\epsilon^{st})
+\sum_{l=0}^{n}\kappa^{(l) \prime}  W D^{(2l)}(\epsilon^{st})\bigr)
W^{-1}\Bigr)\biggr],
\label{deldet}
\end{equation}
where $\epsilon^{2n}$ is the $(2n+1)^{th}$ row of $\epsilon$.
When the infinitesimal variations (\ref{va3}) of the maps are written in
terms of a ghost matrix superfield $\gamma$ (in diagonal basis) instead of
infinitesimal parameters  they become the BRST transformations corresponding
 to the classical super $W$-symmetries. The matrix elements
$\gamma^r_s$, such that $r+s$ is even are nilpotent (i.e.$(\gamma^r_s)^2=
0$) whereas the remaining entries have both a ghost number one and a
grassmannian character. The BRST laws obeyed by the maps ${\cal Z}^i$
are \footnote {As usual the operator $s$ acts as an antiderivation from the
right, the grading being defined by the sum of the ghost number
and the form degree; $s$ does not feel the Grassmann parity.}

\begin{equation}
s {\cal Z}^i =\gamma^i_k {\cal Z}^k-\gamma^{2n}_k {\cal Z}^k
{\cal Z}^i .
\label{br1}
\end{equation}
Nilpotency of the law (\ref{br1}) is fulfilled when

\begin{equation}
s \gamma = -\gamma^2.
\label{br2}
\end{equation}
Therefore this analysis allows to construct with the help of $\gamma$ and $W$
 a matrix $C$ with ghost grading one

\begin{equation}
C =\biggl(\,\sum_{l=0}^{n-1}\kappa^{(l)} \tilde{W} D^{(2l+1)}
(\frac{\hat{\gamma}^{st}}{\Delta})
+\sum_{l=0}^{n}\kappa^{(l) \prime}  W D^{(2l)}
(\frac{\hat{\gamma}^{st}}{\Delta})\biggr)W^{-1},
\label{gmc}
\end{equation}
where

\begin{equation}
\hat{\gamma} =\gamma -
str\Biggl(\biggl(\sum_{l=0}^{n-1}[\kappa^{(l)} \tilde{W} D^{(2l+1)}
 (\gamma^{st})]
+\sum_{l=0}^{n}[\kappa^{(l) \prime}  W D^{(2l)}(\gamma^{st})]\biggr)
W^{-1}\Biggr) \unit .
\label{defg}
\end{equation}
From (\ref{deldet}) and (\ref{br2}) it follows that
\begin{equation}
s({\cal Z}^r \Delta^{-1})= \hat{\gamma}^r_j {\cal Z}^j \Delta^{-1}.
\label{br3}
\end{equation}
From (\ref{br3}) it is then straightforward to deduce the BRST transformation
of the matrix $W$
\begin{equation}
sW=CW.
\label{w1}
\end{equation}

By construction this super-matrix $C$, which is traceless
(as it is straightforward to verify from (\ref{gmc},\ref{defg})),
is independent from map choices in the superprojective structure
${(Z^{i}_{\alpha},\Theta^{i}_{\alpha})}$.
Moreover, under a superconformal coordinate change in
the superholomorphic canonical bundle (given by the transition matrix
$\Lambda$) this superfield transforms as

\begin{equation}
C_{b}= \Lambda_{ba}C_{a} \Lambda_{ba}^{-1}.
\label{w1a}
\end{equation}

The overall consistency of this framework is given by
\begin{equation}
sC=-CC.
\label{w2}
\end{equation}
This law can be proved starting from the expression (\ref{gmc}) of $C$ :
using (\ref{defg}) $C$ can be written
\[
C = B - \sum_{l=0}^{n-1}\kappa^{(l)} D^{2l+1}(\frac{\gamma_D^{st}}{\Delta}) -
\sum_{l=0}^{n}\kappa^{(l) \prime }D^{2l}(\frac{\gamma_D^{st}}{\Delta}),
\]
with
\[
B = \biggl(\sum_{l=0}^{n-1}\kappa^{(l)} \tilde{W} D^{(2l+1)}
 (\frac{{\gamma}^{st}}{\Delta})
+\sum_{l=0}^{n}\kappa^{(l) \prime}  W D^{(2l)}
(\frac{{\gamma}^{st}}{\Delta})\biggr)W^{-1}.
\]
and
\[
\gamma_D = (strB) \unit .
\]
By direct application of $s$ on $B$ with the help of
the explicit
expressions of $\kappa^{(l)}$ and $\kappa^{(l) \prime}$ it is possible to
show that
$$ sB=-C^{2}.$$
From the tracelessness property of $C^{2}$ (which is a result of $strC=0$
and of the ghost grading one of every issue of $C$) it follows that
$s\gamma_{D}=0$ and consequently $sC=-C^{2}$.

From (\ref{w1}) we readily derive the BRST transformation law for
$\om$ and $\os$
\begin{eqnarray}
 s\om & = & DC+\tilde{C}\om-\om C \label{w3} \\
 s\oms & = & \bar{D}C+\tilde{C}\oms-\oms C . \label{w4}
\end{eqnarray}

Now we explain in the following how this matrix formalism allows us to find
the particular BRST algebra which is obeyed by the fields of a given
$SW_{n}$ model.
It is well known that the Ward identities for the induced $W$-gravity are
very similar in structure to the BRST transformations of the projective
connection. On the usual Riemann surface this relation is a straightforward
consequence of the striking similarity between the Beltrami equation
 and the BRST transformation of the projective coordinate $Z$. The same
sort of relations, discussed previously for $W_n$
models in ref.\cite{BG}, are also present in the $SW_n$ models.
 Indeed the comparison of (\ref{holo}) and (\ref{w3}) shows that the
replacement of $(\Db, \oms)$ by $(s,C)$ in (\ref{holo}) leads to
(\ref{w3}) up to some signs \footnote{This sign difference results from the
fact that the operator $D$ acts from the left, whereas $s$ acts from the
right.}.
This allows us to derive the explicit form of $C$ from $\oms$ by replacing
$\mu_{\bar{\theta}}^{\ \theta ...}$ by $c^{\ \theta ...}$,
with the substitution of a
ghost degree to the conformal index $\bar{\theta}$.

From the relations (\ref{w3},\ref{w4}) we compute the BRST laws of
the superfields $\ro{i}$, of the generalized Beltrami
coefficients $\mmu{i}$ and of the superghosts $c_i$.
 From (\ref{w1a}) and the transition laws (\ref{12},\ref{13}) it can be
checked easily that the BRST laws (\ref{w2},\ref{w3},\ref{w4}) are
invariant under a superconformal coordinate change. Thus they are well
defined on the SRS.

In summary the laws (\ref{w2},\ref{w3},\ref{w4}) represent the nilpotent BRST
 algebra (as it can be verified by an explicit calculation) corresponding to
a given classical $SW$-algebra. They are obtained
thanks to the definition (\ref{gmc}) which induces the BRST transformations
(\ref{w1},\ref{w2}), once the law (\ref{br2}) has been chosen.

\subsection{Super covariant anomalies}

Using the coboundary operator $d$ defined as
$$ d \Phi = D\Phi d\sz	+ \Db \Phi d\szb $$
and the laws (\ref{w3},\ref{w4}) we can write the transformation of the
connection $\con$ defined in (\ref{scon}) :
\begin{eqnarray}
s \con & = &- d C - \Ct {\con} - {\con} \Ct \label{a1}	 \\
s \tilde{\con} & = &- d \Ct - C \tilde{\con} - \tilde{\con} C \label{a1bis}
\end{eqnarray}
where $ \widetilde{d C} = d \Ct$ results from a cancellation
of minus signs between the derivative $\widetilde{DC} = -D\Ct$ and
$\widetilde{d\sz} = -d\sz$.
The  polynomial $T_3^0$ of rank 3 (where the lower index denotes
the form degree and the upper index the ghost number)

     $$T_3^0=str(\con D \con +\frac{2}{3} \con \con \con) . $$

\noindent generates a tower of	descent equations through the application
of the
BRST transformations (\ref{w2},\ref{a1},\ref{a1bis})
\begin{eqnarray}
sT_3^0+dT_2^1&=&0      \nonumber   \\
sT_2^1+dT_1^2&=&0      \label{d1}  \\
sT_1^2+dT_0^3&=&0      \label{d2}  \\
sT_0^3=0,	      \label{d3}
\end{eqnarray}

\noindent where the explicit expressions of the cocycles are given by

\begin{eqnarray}
T_2^1& = &-str(\Ct\om \omst + \Ct\os \omt)d\sz d\szb  \label{f1} \\
T_1^2& = &str(C^2\omt) d\sz+ str(C^2\omst) d\szb   \label{f2}  \\
T_0^3& = &-\rat{1}{3}str(\Ct^3) .     \label{f3}
\end{eqnarray}
 Eq.(\ref{d1}) implies $s\int T_2^1=0$ and identifies this descendant as a
candidate for a consistent anomaly, a non-trivial solution of the
Wess-Zumino \cite{WZ} consistency condition.  \\
The anomalous cocycle $T_2^1$ does not transform tensorially,
as can be
easily verified by using eqns. (\ref{12},\ref{13},\ref{w1a}).
However  it can be written, using (\ref{holo}),
\[
T_2^1 = str( \Ct\Db \omt+ \Ct D \omst)d\sz d\szb .
\]
As one can straightforwardly verify, only the first term of the sum
\begin{equation}
\sig{2}{1} = str(\Ct\Db \omt) \label{sig21}
\end{equation}
is well-defined on a SRS. It also solves the descent equations
(\ref{d1}-\ref{d3}) and is thus a candidate for a covariant and
consistent anomaly \footnote{Covariant expressions for the
anomaly have already been obtained in \cite{gierop} and for the
bosonic case in \cite{zucchini,BG,grimm}.}.
The corresponding cocycles are
\begin{eqnarray}
\sig{1}{2} & = & \rat{1}{2} str(CD\Ct + 2C\omt\Ct)d\sz +
\rat{1}{2}str(\Db\Ct\Ct)d\szb \label{sig12}\\
\sig{0}{3} & = & \rat{1}{6} str(\Ct^3) .   \label{sig03}
\end{eqnarray}
The fact that $\con $ is not a generic connection  since the matrix
elements of $\om$ and $\oms$ are not all independent but linked
by eq.(\ref{holo}) plays no role here (C is not constrained since, as
mentionned  before, (\ref{holo}) becomes (\ref{w3}), the transformation
law of $\om$, after suitable substitutions). Actually, the nilpotency
of $s$, which is a crucial ingredient in this framework, is independent
of these constraints, as it can be straightforwardly ascertained from
(\ref{w2},\ref{w3},\ref{w4}). At last the residual conditions given by
(\ref{holo}) after the determination of some entries of $\oms$, are
the super holomorphy conditions obeyed by the $\rt{i}$ ; they serve,
as explained above, to relate the two sets (\ref{f1},\ref{f2},\ref{f3}) and
(\ref{sig21},\ref{sig12},\ref{sig03}) of cocycles .

The formalism presented above provides us
with a completely algorithmic procedure of calculating the anomaly
associated to a given super $W$-model and the cocycles related to this
anomaly by the system of descent equations. Moreover the solution
(\ref{sig21},\ref{sig12},\ref{sig03}) has the advantage of being defined
on a generic SRS of arbitrary genus. The form of the Virasoro Ward
identity on an arbitrary Riemann surface was first derived in \cite{LS};
however there an holomorphic projective connection ${\cal R}$, which was
BRST inert, was introduced by hand. In contrast, the formulation presented
here has the advantage of being self-contained since it is the usual
projective
connection that renders the local expression of the super-anomaly
well-defined.

 Since we do not start as usual from an action, but construct in an
algebraic way expressions of the BRST anomaly and of its cocycles which obey
to the descent equations, let us discuss more precisely the kind of anomalies we
obtain. The so-called universal $W$-gravity anomalies (for a review see
ref.\cite{r2}) that are present in all
 theories of matter coupled to W-gravity, are those anomalies that depend
only on the gauge fields $\tilde{\mu_{i}}$ and not on the matter fields. For
theories in which the symmetry is non-linearly realized the universal form of
 the anomaly in the spin $s$ symmetry (by reference to the spin of the
corresponding current $\tilde{\rho_{i}}$) is given by $\tilde{\mu_{i}}
\drp^{s+1} c^i$ where  $c^i$ denotes the ghost associated to this symmetry.
In a framework where a special realization of the currents in terms of scalar
 fields is considered, these anomalies arise at $s-1$ loops level. At lower
number of loops there are anomalies which depend on matter fields. The
supersymmetric extension of these universal anomalies is the subject of this
chapter, where the equivalent of the corresponding universal expression
given above is dressed with $\tilde{\rho_{i}}$ dependent terms in order to
insure the BRST invariance for consistent anomalies and both BRST
invariance and conformally covariance for covariant anomalies.

\section{Supergravity}

This section is an illustration of the general formulation presented here
for the
particular case $n=1$ (i.e. $3\times 3$ matrices). This corresponds
to the SRS approach of the $(1,1)$ supersymmetry.
First, the model is studied in its full generality and then the restricted
geometry given by the W-Z gauge is discussed thus making contact with
the usual supergravity and the \sv\  algebra . Finally advantages
with respect to previous approaches are emphasized.

\subsection{The underlying classical \sv\  algebra}

We begin with a connection form ${\cal{A}}$ built from the matrix
$W$ through the definitions
(\ref{om},\ref{omstar}). The two components of this connection $1$-form are,
in the block grading,
$$
\om=\left( \begin{array}{ccc}
		   0 & 0 & 1 \\
		   \rho_{3} & 0 & \rho_{2} \\
		   0 & 1 & 0
		  \end{array}
	   \right),
    \  \  \  \	\  \  \
    \os=\left( \begin{array}{ccc}
		   \alpha_{1} & \mu_{3}     & a_{1} \\
		   \alpha_{2} & \alpha_{3}  & a_{2} \\
		   a_{3} & \mu_{2}     & \alpha_{4}
			 \end{array}
		  \right),
$$
and equivalently in the diagonal grading
\[ \om=\left( \begin{array}{ccc}
		   0 & \rho_{2} & \rho_{3} \\
		   1 & 0 & 0 \\
		   0 & 1 & 0
		  \end{array}
	   \right),
    \  \  \  \	\  \  \
    \os=\left( \begin{array}{ccc}
		   \alpha_{3} & a_{2} & \alpha_{2} \\
		   \mu_{2} & \alpha_{4}  & a_{3} \\
		   \mu_{3} & a_{1} & \alpha_{1}
			 \end{array}
		  \right).
\]
The conformal indices of the four independent fields are
$$ \rho_{2}\mid_{z} \ \ \ \rho_{3}\mid_{z\theta} \ \ \
\mu_{2}\mid_{\tbar}^{\ \theta} \ \ \ \mu_{3}\mid_{\tbar}^{\ z}.$$
Building explicitly the matrix $\Lambda$ and using relations (\ref{12})
and (\ref{13}),
we determine the variation of the coefficients under a superconformal change
of coordinates $\sz\longrightarrow \sz'$ (we recall that $e^{-X}=D\theta'$)
\begin{eqnarray}
  \rho_{3}' & = & e^{3X}\left[\rho_{3}+S(z,\theta;\theta^{\prime})
			    +(DX)\rho_{2} \right] \label{conf1}\\
  \rho_{2}' & = & e^{2X}\rho_{2}  \label{conf2} \\
	    &  &  \nonumber \\
  \mu_{3}' & = & e^{\bar{X}}e^{-2X}\mu_{3} \\
  \mu_{2}' & = & e^{\bar{X}}e^{-X}(\mu_{2}+(DX)\mu_{3})
\end{eqnarray}
where $S(z,\theta;\theta^{\prime})$ is the super-Schwarzian derivative of
this superconformal change of coordinates \cite{friedan}.
Then if $\rho_{2}=0$,  $\rho_{3}$ transforms as a
superprojective connection.

The matrix $W$ considered as an element of $sl(2\!\mid\!1)$ can
be parametrized in the diagonal basis using a Gauss decomposition

\begin{equation}
\label{Gauss}
W=\left( \begin{array}{ccc}
		   1 & 0 & 0 \\
		    \phi_{2} & 1 & 0 \\
		   f &	\phi_1 & 1
		  \end{array}
	   \right)
    \left( \begin{array}{ccc}
		   \lambda_{1}^{-1}\lambda_{2}^{-1} & 0  & 0  \\
			  0	& \lambda_{2}^{-1}  & 0  \\
		     0	 &   0	 &  \lambda_{1}
	 \end{array}
		  \right)
\left( \begin{array}{ccc}
		     1	 &  \Phi_{1} & F \\
		    0 & 1  &  \Phi_{2}	\\
		    0 & 0   & 1
			 \end{array}
		  \right), \label{gauss}
\end{equation}
where the elements $\Phi_{i}$, and $\phi_{i}$ are Grassmann variables.
Comparing with (\ref{defw}) we can express all the variables in terms
of the superprojective coordinates $Z$ and $\Th$.
They are given by:\\[5mm]
\begin{tabular}{rclrclrcl}
$\phi_1 $& = & $Dln \lambda_1 $, &\hspace{15mm}$\lambda_1$ & = &
$(D\Th)$, & \hspace{15mm}$\Phi_1 $& = & $\Th$,\\[5mm]
$\phi_2 $& = &$-Dln \Delta$, &\hspace{15mm}$\lambda_2$ & = &
$D(\frac{DZ}{D\Th})(D\Th)^{-1}$,&
\hspace{15mm}$\Phi_2$ & = & $ (DZ) \lambda_1^{-1}$,  \\[5mm]
$f$ & = & $-\partial ln \Delta$,&  & & &
\hspace{15mm}$F $& = & $Z,$
\end{tabular}\\[5mm]
where $\Delta$ is
\begin{equation}
\Delta=D(\frac{DZ}{D\Th}).
\end{equation}

The entries of the $\om$ matrix $\alpha_{i}$ and  $a_{i}$ are determined
by solving the system given by (\ref{holo}) when $\Db \om$ has a
null contribution.
The remaining equations of this system
are the holomorphy conditions for $\rho_{3}$ and $\rho_{2}$. They
take the forms
\begin{equation}
\label{ward}
\begin{array}{rcl}
   \Db\rho_{3}-\mu_{3}\partial\rho_{3}-\mu_{2}D\rho_{3}-(2\partial
\mu_{3}+D\mu_{2})\rho_{3}+\partial^{2}\mu_{2}-\rho_{2}\partial\mu_{2} &=&0,
 \\
   \Db\rho_{2}-\partial^{2}\mu_{3}-2D\partial\mu_{2}-\partial
(\mu_{3}\rho_{2})+\rho_{3}D\mu_{3}-\mu_{2}D\rho_{2}+2\rho_{3}\mu_{2}
& = & 0.
\end{array}
\end{equation}
To exploit these relations we have to define the physical fields and first
to give a physical meaning to $\rho_3$, i.e. to turn this field into a
superprojective connection. From the expressions
(\ref{conf1}) and (\ref{conf2})  it
is easy to guess the right term to add to $\rho_3$. Of course
 the Drinfeld-Sokolov method described in sect. 3.1  leads to the
same result which is
\begin{equation}
\begin{array}{lll}
\rt{2}&=&\f{1}\rho_{2}, \\
 \rt{3}&=&\rho_{3}-\f{1}D\rho_{2}.
\end{array}
\label{rt}
\end{equation}
To covariantize $\mu_{3}$
 we use the method of the supersymplectic form (see sect. 3.2) by
considering :
$$ \omega=\int_{\Sigma}(\delta\mu_{3}\wedge\delta\rho_{3}
+\delta\mu_{2}\wedge\delta\rho_{2}),$$
and replacing the $\rho_{k}$ by the $\rt{k}$:
\begin{eqnarray*}
 \omega & = & \int_{\Sigma}(\delta\mu_{3}\wedge\delta\rt{3}
+\delta\mu_{3}\wedge \delta D\rt{2}+\delta\mu_{2}\wedge\delta(2\rt{2})) \\
	& = & \int_{\Sigma}(\delta\mu_{3}\wedge\delta\rt{3}
   +\delta(2\mu_{2}+D\mu_{3})\wedge\delta\rt{2}).
\end{eqnarray*}
We thus obtain :
\begin{equation}\begin{array}{lll} \mt{3}&=&\mu_{3} \\
 \mt{2}&=&2\mu_{2}+D\mu_{3} \end{array}\label{mt}
\end{equation}
and as one can directly verify these new fields transform tensorially.

Let us note that the Gauss decomposition induces for these fields
 surprisingly simple expressions

\begin{eqnarray}
\rt{2}&=&-\frac{1}{2}\partial ln\lambda_2 + \frac{1}{2}Dln \lambda_1
Dln \lambda_2, \label{rho2}\\
\rt{3}&=&-S(z,\theta;\Th)-\frac{1}{2}\partial Dln\lambda_2
+\frac{1}{2}Dln \lambda_1 \partial ln \lambda_2
 +\frac{1}{2}Dln \lambda_2 \partial ln \lambda_1.
\label{rho3}
\end{eqnarray}
These expressions involve only the parameters entering in the central
matrix of the Gauss decomposition (\ref{gauss}).
On an ordinary SRS, from the $\frac{1}{2}$-super differential
$\lambda_1$ is built
the super-affine connection $\zeta=Dln\lambda_1$ which, in turn allows
one to define the super-Schwarzian derivative :

\begin{equation}
S(z,\theta;\Th)=\partial \zeta-\zeta D\zeta. \label{S}
\end{equation}
In this kind of generalization of the SRS, the partner $ \lambda_2$
of $\lambda_1$	appears in (\ref{rho2},\ref{rho3}) in expressions
 very reminiscent of the usual super-affine connection (\ref{S}). However
the terms in the right hand side of (\ref{rho3}) (except $S$ of course) and
\rt{2} transform covariantly.

Now we replace the $\rho_{k}$'s and the $\mu_{k}$'s by the corresponding
$\rt{k}$'s and $\mt{k}$'s in the identities (\ref{ward}) and obtain
\begin{equation}
 \Db\rt{3} = \f{1}\bol{2}{-\rt{3}}\mt{3}+(\rt{2}\partial
-\f{1}D\rt{2}D+\f{1}\partial\rt{2})\mt{2},
\label{21a}
\end{equation}
\begin{equation}
 \Db\rt{2}
	     =\f{1}\bol{1}{-\rt{3}}\mt{2}+(\rt{2}\partial-\f{1}D\rt{2}D
+ \drp\rt{2})\mt{3},	\label{21b}
\end{equation}
where ${\cal L}_1(-\rt{3})$ and ${\cal L}_2(-\rt{3})$ are the super Bol
operators \cite{gierop} of the superprojective connection ${\cal{R}}=-\rt{3}$
\beqnas
{\cal L}_1({\cal R}) & = &  D^3 + {\cal R}, \\
{\cal L}_2({\cal R}) & = &  D^5 + 3{\cal R}D^2 + (D{\cal R})D
+ 2(D^2{\cal R}).
\eeqnas
The algebraic content of these holomorphy equations  is given
by the Poisson brackets among the spin-one superfield
$\rt{2}$
and the superfield of spin $3/2$ $\rt{3}$.
It corresponds to the $N=2$ classical super Virasoro algebra
 \cite{huitu}.
If we further impose that the connection $1-$form be an \osp\ connection
instead of
an $sl(2\!\mid\!1)$ one, the fields $\rt{2}$ and $\mt{2}$ become zero.
By an explicit
calculation of $\rho_{2}$ in terms of the coordinates $Z$ and $\Th$ we
found that setting $DZ=\Th D\Th$ brings $\rho_{2}$ to zero and $\rho_{3}$
to $-S$.

\subsection{BRST analysis and the covariant anomaly}

The BRST transformations of the superprojective coordinates are

\begin{eqnarray}
sZ &=& \ct{3} \drp Z+\f{1} D \ct{3 }DZ- \f{1} \ct{2} DZ  \label{s23}\\
s \Th &=& \f{1} D \ct{3} D \Th +\ct{3} \drp \Th - \f{1} \ct{2} D\Th
\label{s24}
\end{eqnarray}

They are given in terms of the covariant (tilde) fields $\ct{k}$ whose
expressions as functions of the $c_i$  are similar to ($\ref{mt}$).
\begin{equation}
\ct{3}=c_3	 ,\qquad	\ct{2}=2c_{2}-Dc_{3}.
\end{equation}
Then the construction of the BRST algebra follows from (\ref{w2}-\ref{w4})
which insure its completeness and its nilpotency. It is given
here in terms of the covariant (tilde) fields $\rt{k},\  \mt{k}$, defined
by (\ref{rt},\ref{mt}).
We obtain
\begin{eqnarray}
s \ct{3} & = & -\ct{3}\partial\ct{3}-\frac{1}{4}D\ct{3}D\ct{3}
		+\frac{1}{4}\ct{2}\ct{2},	 \label{sa1}   \\
s \ct{2} & = & -\ct{3}\partial\ct{2}-\frac{1}{2}\ct{2}\partial\ct{3}
	       -\frac{1}{2} D\ct{3} D\ct{2}.
\label{sa2}
\end{eqnarray}
The fields $\rt{2},\ \rt{3}$ obey the	     following laws
\begin{eqnarray}
s \rt{3} & = & -\f{1}\bol{2}{-\rt{3}}\ct{3}+(\rt{2}\partial
-\f{1}D\rt{2}D+\f{1}\partial\rt{2})\ct{2}, \label{s7}	\\
s \rt{2} & = & -\f{1}\bol{1}{-\rt{3}}\ct{2}+(\rt{2}\partial
-\f{1}D\rt{2}D+\partial\rt{2})\ct{3},
\label{s8}
\end{eqnarray}
and the $\mt{2},\ \mt{3}$ fields satisfy
\begin{eqnarray}
s \mt{3} & = & \Db\ct{3}+\ct{3}\drp\mt{3}-\mt{3}\drp\ct{3}
		 +\f{1}D\mt{3}D\ct{3}+\f{1}\ct{2}\mt{2}, \label{s9}   \\
s \mt{2} & = & \Db\ct{2}-\f{1}\mt{2}\drp\ct{3}+\f{1}D\ct{3}D\mt{2}
		+\ct{3}\drp\mt{2}-\f{1}\ct{2}\drp\mt{3}   \nonumber \\
	    &	  &   +\f{1}D\mt{3}D\ct{2}-\mt{3}\drp\ct{2}.
\label{s10}
\end{eqnarray}
Having at hand the BRST relations for all the fields, we can address the
problem of the anomaly. Using relation (\ref{sig21}) we can write down
explicitly
\begin{eqnarray}
\sig{2}{1} & = &
\{\ct{2}(-{\cal L}_1(-\rt{3})\mt{2}+D\mt{3}D\rt{2}-2\rt{2}D^{2}\mt{3}-
2\mt{3}D^{2}\rt{2})+ \nonumber \\
 & &\ct{3}({\cal L}_2(-\rt{3})\mt{3} -D\rt{2}D\mt{2}+
\mt{2}D^{2}\rt{2}+2\rt{2}D^{2}\mt{2})\}d\sz d\szb   \label{sig21expl}.
\end{eqnarray}
It is worth noting that by symmetrizing this covariant anomaly we get the
compact  form $\displaystyle{\sig{2}{1} = \sum_{i=2}^3 (\mt{i} s \rt{i} -
\ct{i} \Db\rt{i})}d\sz d\szb$.

\subsection{Comparison with previous works}

Super-Beltrami differentials were previously introduced either by using
zweibeins \cite{BBG} or with the help of super 1-forms \cite{dgbel}.
In this latter approach super-Beltrami differentials occur without any
reference to metrics or vielbeins thanks to the super 1-forms $e^Z \equiv dZ
+ \Th d\Th$ and $e^\Th \equiv d\Th$ ( and c.c.) which span the
cotangent space of the SRS. Their expressions with respect to a reference
coordinate system yields six superfields ($H_{\zbar}^{\ z},
H_{\theta}^{\ z},H_{\tbar}^{\ z},H_{\theta}^{\ \theta},
H_{\zbar}^{\ \theta}, H_{\tbar}^{\ \theta}$). The structure equations
$d e^Z + \Th d \Th = 0 = d e^{\Th}$ (and c.c.) relate four of these
superfields to only two independent Beltrami coefficients
$H_{\theta}^{\ z}, \bel$.
In our parametrization we also obtain two independent fields
$\mt{2}$ and $\mt{3}$, which could be compared, due to their conformal
weights, to $\bel$ and $H_{\tbar}^{\ \theta}$, namely
\begin{displaymath}
\bel =\mt{3}+\left( DZ-\Th D\Th \right)
\frac{D\Th\partial\Th\Db Z}{(\partial Z)^{2}D\Th}
\end{displaymath}
\begin{displaymath}
H_{\tbar}^{\ \theta}=-\f{1}\frac{D\Th}{\sqrt{\partial Z
+\Th\partial\Th}}(\mt{2}+D\mt{3})
-\left( DZ-\Th D\Th \right)\frac{\Db\Th\partial\Th}{(\partial Z)^{3/2}D\Th}.
\end{displaymath}
Whereas our fields transform
homogeneously under a superconformal change of the coordinate system (more
precisely they are sections $e^{\bar{X}}e^{-X}$ and $e^{\bar{X}}
e^{-2X}$, respectively, of the canonical fibre bundle), the transformation
laws of $\bel$ and $H_{\tbar}^{\ \theta}$ do not take simple forms.
They depend in particular on the superfield $H_{\theta}^{\ z}$.
Furthermore, our fields $\mt{2}$
and $\mt{3}$ appear in the context of $sl(2\mid \! 1)$, while, in contrast,
it is possible to obtain both $H_{\theta}^{\ z}$ and $\bel$ in the context
of \osp\ \cite{FD}.

The choice made in sect.5.2
\footnote{We can use the formula (\ref{gmc}) to express the superghosts
 $\ct{3}$,
$\ct{2}$ which are the entries $<20>$ and $<10>$ respectively of the
matrix $C$ in terms of the superghosts $\gamma$
\begin{eqnarray*}
\ct{3} & = & (\gamma_1+2\gamma_0 Z -\gamma_2 Z^{2} - \gamma_e \Th -\gamma_d
\Th Z) \frac{1}{\drp Z}(1-\frac{DZ \drp \Th}{\drp Z D \Th})   \nonumber  \\
  & - &(\gamma_d Z + \gamma_e +\gamma_0 \Th -\gamma_2 \Th Z)\frac{DZ}{\drp
Z D \Th},   \label{s19} \\
\ct{2} & = & -D c_3 -(\gamma_d Z + \gamma_e +\gamma_0 \Th  -\gamma_2 \Th Z)
\frac{1}{D \Th}(1+\frac{DZ \drp \Th}{\drp Z D \Th})   \nonumber  \\
&+&(\gamma_1 +2 \gamma_0 Z -\gamma_2 Z^{2} - \gamma_e \Th -\gamma_d \Th Z)
\frac{\drp \Th}{\drp Z D \Th},
\end{eqnarray*}
where we have assumed for simplicity that the $\gamma$'s are antiholomorphic.}
 for the two
independent superghosts $c_3$ and $c_2$ is not unique. Indeed,
if we take as generators of the BRST transformations the superghost fields
$c^z$, $c^{\theta}$ defined by
\begin{eqnarray*}
C^Z & = & c^z\lzz \\
C^{\Th} & = & c^{\theta}\sqrt{\lzz}+c^z\partial \Th
\end{eqnarray*}
where $\lzz=\partial Z+\Th\partial\Th$ and where $C^Z$ and $C^{\Th}$ are
\begin{eqnarray*}
C^Z & = & \gamma_1+2\gamma_0 Z-\gamma_2 Z^2 \\
C^{\Th} & = & \gamma_d Z+\gamma_e+\gamma_0 \Th-\gamma_2 \Th Z
\end{eqnarray*}
we obtain instead of (\ref{s23},\ref{s24}) the BRST laws given in
\cite{dgbel}. This remark completes the comparison between our
formalism and this work; both approaches
start from the same gauge transformations, namely the relations
(\ref{13a},\ref{14a}) but then differ by the choice of the infinitesimal
parameters which turned into ghost fields become the generators of the BRST
transformations.

 The well-known results of the $N=1$ SRS \cite{dgbel,Takama}
restricted to the W-Z gauge, where $H_{\tbar}^{\ z}=0$
($\mt{2}=0$), are easily reproduced. In this
case the pair of geometrical superfields usually encountered in the
literature, namely the super Schwarzian derivative $S(\Theta;z,\bar{z}
,\theta,\bar{\theta})$ \cite{friedan} and the super Beltrami superfield
$H^{\ z}_{\bar{\theta}}$ \cite{dgbel} correspond respectively to
$\tilde{\rho}_3$ and $\mt{3}$. Moreover, the holomorphy condition for
$\rt{3}$, which is the superconformal anomalous Ward identity obtained by
replacing
$\tilde{\rho}_3 \to \frac{\delta \Gamma}{\delta \mu}$ (where $\Gamma$ is the
generating functional for current correlation functions), is recovered and
appears, as expected, as a compatibility condition between  $\rt{3}$ and
$\mt{3}$ following from (\ref{21a}) by putting $\mt{2}=0$. Otherwise the
superghost $C^z$ is given by $\ct{3}$, the laws (\ref{sa1},\ref{s7},\ref{s9})
reduce to the well known transformations \cite{OSSV,dgano} and it
is not very hard to check that (\ref{sig21expl}) gives the standard
super-diffeomorphism anomaly \cite{OSSV,dgbel}.

We finally briefly discuss the $SW_{3}$ example ($n=2$) by comparing it with
existing results \cite{theisen}.
Previous studies of this case are few and uncomplete. However
in \cite{theisen} the link between the $N=1$ super $W$-algebra and
$osp(2m\pm 1\!\!\mid\!\! 2m)$ was studied. The authors developed the example of
$osp(3\!\!\mid\!\! 2)$ and gave the Poisson brackets  among fields $V_i$'s
which are
related to our fields $\rt{i}$'s through the correspondence
$V_i \llra - \rt{i} ,\	 3H_i \llra \mt{i}$. A more general treatment of
this example is given elsewhere \cite{abn}.
\section{Discussion and outlook}
Let us conclude on some future prospects.
Amazingly the theory presented here appears to yield new bosonic models.
In the conventional geometrical framework \cite{zucchini,OSSV}, the  spins
of the generators of the algebra are limited to ($2,3,\ldots,n$). In the
present formalism the spin content of the supergenerators is given by
($(1,\rat{3}{2}),(\rat{3}{2},2),(2,\rat{5}{2})\ldots,(\f{2n-1},n)$)
where the decomposition in components has been made explicit. The limit of
purely bosonic generators $(1,2,2,3,\ldots,n)$ which is
obtained when the expansion in component fields is limited to the scalar
term  for even spin generators and to the $\theta$ term for odd spin
generators, supplies a new spin 1 current and duplicates the standard
currents. Thus such a framework interestingly enough allows one to construct
a spin 1 field which cannot be obtained in the conventional bosonic
approach. Hence, for the first time, to our knowledge, a  bosonic limit of a
supersymmetric framework is obtained which is impossible to get in the
standard bosonic scheme.

Moreover it is	alluring to study these supersymmetric models since,
compared to the
underlying bosonic theories, they contain a rich gauge choice, generalizing
the W-Z gauge. Particular subsets of fields are selected by setting
to zero some
superfields \rt{i}. Then the holomorphy relations obeyed by these
superfields
become constraints for the super-Beltrami differentials. These constraints are
more or less tractable. Two situations can occur.\\
Thanks to the constraints it might be straightforward to eliminate some
of the super-Beltrami differentials as explicit functions of the remaining
ones.
In general these relations can be associated to some group prescription by
assuming that the matrix $\Omega$ belongs to some representation of the Lie
algebra of a super group. This is the case for instance for the choice made
in \cite{theisen} for $SW_3$. \\
More difficult is the situation where the constraints appear
as differential equations implying an implicit dependence of some Beltrami
differentials on other ones. In that case  it seems always possible to extract
the classical super algebras corresponding to this choice, without being able
to write the corresponding Ward identities and anomaly. This indicates that
these algebras cannot be used in a construction of some supersymmetric
generalization of string theory and thus, gives a criterion to determine if the
classical super algebra has a physical meaning or not.
~

~

\noindent{\Large{\bf Acknowledgments}}
\vspace{0.8cm}
\newline
J.P.A. acknowledges M. Abud and L. Capiello for discussions.
We thank J. T. Donohue for a careful reading of the manuscript
and F. Delduc and F. Gieres for many criticisms and useful
comments on a preliminary version of this work.
Y.N. is grateful to the University of Bordeaux I for granting him a discharge
during a part of this work.

\section{Appendix}

For \sln\ we take the following Cartan matrix \cite{noh}
$$ C_{i,j}=(-1)^{i+1}\delta_{i+1,j}+(-1)^{i}\delta_{i,j+1} .$$
The Chevalley basis in the diagonal grading is given by
\begin{eqnarray*}
	h_{i} & = & (-1)^{i+1}\left( E_{i,i}+E_{i+1,i+1} \right) \\
	e_{i} & = & (-1)^{i+1} E_{i,i+1} \\
	f_{i} & = & E_{i+1,i}
\end{eqnarray*}
where $(E_{ij})_{kl}=\delta_{ik}\delta_{jl}$.
We have the relations
\begin{eqnarray*}
\left[ h_{i},e_{j} \right]  & = & C_{ij}e_{j} \\
\left[ h_{i},f_{j} \right]  & = & -C_{ij}f_{j} \\
\left[ e_{i},f_{j} \right]  & = & \delta_{ij}h_{i} .
\end{eqnarray*}
To construct the \osp\ subalgebra generating matrices we need the inverse
of the Cartan matrix. It is given by
\begin{eqnarray*}
C^{-1}_{2p,j} & = & \sum_{k=0}^{p-1} \delta_{2k+1,j} \ \ \   1\leq p\leq n \\
C^{-1}_{2p+1,j} & = & \sum_{k=p+1}^{n} \delta_{2k,j} \ \ \   0\leq p\leq n-1.
\end{eqnarray*}
Thus
\begin{eqnarray*}
	J_{+} & = & \sum_{i} \left( ie_{2i}+(n-i)e_{2i+1} \right) \\
	H     & = & \sum_{i} \left( ih_{2i}+(n-i)h_{2i+1} \right) .
\end{eqnarray*}
As shown in \cite{theisen} it is possible to find a Chevalley basis for
$osp(2m\pm1\!\mid\!2m)$ such that the generators $J_{\pm}$ and $H$
of the \osp\ subalgebra have the same expressions as for \sln.
For $osp(2m+1\!\mid\!2m)$ it is given by
\begin{eqnarray*}
h_{i} & = &
(-1)^{i}\left( E_{2m+1-i,2m+1-i}-E_{2m+1+i,2m+1+i}+E_{2m+2-i,2m+2-i}-
E_{2m+i,2m+i}\right) \\
e_{i} & = & (-1)^{i}\left( E_{2m+i,2m+1+i}-E_{2m+1-i,2m+2-i} \right) \\
f_{i} & = &  E_{2m+1+i,2m+i}+E_{2m+2-i,2m+1-i}
\end{eqnarray*}
and for $osp(2m-1\!\mid\!2m)$ by
\begin{eqnarray*}
h_{i} & = &
(-1)^{i+1}\left( E_{2m-i,2m-i}-E_{2m+i,2m+i}+E_{2m+1-i,2m+1-i}-E_{2m-1+i,2m-
1+i}\right) \\
e_{i} & = & (-1)^{i}\left( E_{2m-i,2m+i}-E_{2m-i,2m+1-i} \right) \\
f_{i} & = &  E_{2m+i,2m-1+i}+E_{2m+1-i,2m-i}.
\end{eqnarray*}


\begin{thebibliography}{50}
\bibitem{r1} P. Bouwknegt and K. Schoutens, \rep{223}{1993}{183}.

\bibitem{r2} C.M. Hull, Lectures on W-Gravity, W-Geometry and W-Strings,
Trieste Summer School in High Energy Physics and Cosmology (1992).

\bibitem{SSZ}
G.M. Sotkov and M. Stanishkov, C.-J. Zhu, \np{356}{1991}{245};
\newline G.M. Sotkov, M. Stanishkov, \np{356}{1991}{439}.

\bibitem{zucchini} R. Zucchini, \cqg{10}{1993}{253}.

\bibitem{gerv} J.-L. Gervais, Introduction to differential W-Geometry,
Lectures given at the workshop Strings, Conformal models and Topological
Field Theories, Carg\`ese (1993) and the meeting "String 93", Berkeley.
\newline J.-L. Gervais and Y. Matsuo, \pl{274}{1992}{309},
\cmp{152}{1993}{317}.

\bibitem{Pol} A.M. Polyakov, \mpl{2}{1987}{893}.

\bibitem{pw} A. M. Polyakov and P. Wiegemann, \pl{131}{1983}{121}.

\bibitem{OSSV} H. Ooguri, K. Schoutens, A. Sevrin and
P. van Nieuwenhuizen, \cmp{145}{1992}{515}.

\bibitem{theisen} F. Gieres and S. Theisen, \ijmp{9}{1994}{383}.

\bibitem{ds} V.G. Drinfeld and V.V. Sokolov,
J. Sov. Math. {\bf30} (1985) 1975.

\bibitem{CR} L. Crane and J.M. Rabin, \cmp{113}{1988}{601}.

\bibitem{dgbel} F. Delduc and F. Gieres, \cqg{7}{1990}{1907}.

\bibitem{Takama} M. Takama, \cmp{143}{1991}{149}.

\bibitem{gt} F. Gieres and S. Theisen, \jmp{34}{1993}{5964}.

\bibitem{Pol2} A.M. Polyakov, \mpl{5}{1990}{833}.

\bibitem{bfk} A. Bilal, V.V. Fock and I.I. Kogan, \np{359}{1991}{635}.

\bibitem{Das} A. Das, W.-J. Huang and S. Roy, \ijmp{7}{1992}{3447}.

\bibitem{BG} J. de Boer and J. Goeree, \np{401}{1993}{369}.

\bibitem{zuber} P. Di Francesco, C. Itzykson and J.-B. Zuber,
\cmp{140}{1991}{543};
   \newline  M. Bauer, P. Di Francesco, C. Itzykson and J.-B. Zuber,
\np{362}{1991}{515}.

\bibitem{bersh} M. Bershadsky, W. Lerche, D. Nemeschansky and N.P. Warner,
 \np{401}{1992}{304}.

\bibitem{sevrin} A. Sevrin, K. Thielemans and W. Troost, \np{407}{1993}{459}.

\bibitem{noh} H. Nohara and K. Mohri, \np{349}{1991}{253}
     \newline S. Komata, K. Mohri and H. Nohara, \np{359}{1991}{168}.

\bibitem{frs} L. Frappat, E. Ragoucy and P. Sorba, \cmp{157}{1993}{499}.

\bibitem{nahm} M. Scheunert, W. Nahm and V. Rittenberg, \jmp{18}{1977}{155}.

\bibitem{wipf} J. Balog, L. Feh\'er, P. Forg\'acs, L. O'Raifeartaigh and A.
Wipf,
\pl{227}{1989}{214}, \pl{244}{1990}{435}, Ann. Phys. {\bf 203} (1990) 76.

\bibitem{gierop} F. Gieres, \ijmp{8}{1993}{1}.

\bibitem{fk} V.V. Fock and I.I. Kogan, \mpl{5}{1990}{1365}.

\bibitem{boo} M. Bershadsky and H. Ooguri, \cmp{126}{1989}{49},
\pl{229}{1989}{375}.



\bibitem{drs} F. Delduc, E. Ragoucy and P. Sorba, \cmp{146}{1992}{403}.

\bibitem{Pope}C.N.  Pope, {\em W-strings 93}, Spring Workshop on High-
Energy Physics, Trieste, April 1993; {\em Strings 93}, Berkeley, May 1993.

\bibitem{MPSX}R. Mohayee, C.N. Pope, K.S. Stelle and K.-W. Xu,
\np{433}{1995}{712}.

\bibitem{Nin} H. Ninnemann, \cmp{150}{1992}{267}.

\bibitem{WZ} W. Bardeen and B. Zumino, \np{244}{1984}{421}.

\bibitem{grimm} D. Garajeu, R. Grimm and S. Lazzarini, CPT 94/P3078 and
hep-th/9411125.

\bibitem{LS} S. Lazzarini and R. Stora, in "knots, topology and quantum
field theories", Proc. 13th Johns Hopkins Workshop, eds Lusanna
 (World Scientific, Singapore, 1989).

\bibitem{friedan} D. Friedan in {\em Unified String Theories}
(Santa Barbara Workshop), eds. M.B. Green , D.J. Gross
(World Scientific, Singapore,1986).

\bibitem{huitu} K. Huitu and D. Nemeschansky, \mpl{6}{1991}{3179}.


\bibitem{BBG} L. Baulieu, M. Bellon and R. Grimm, \np{271}{1989}{697}.

\bibitem{FD} F. Delduc, private communication.

\bibitem{dgano} F. Delduc and F. Gieres, \ijmp{7}{1992}{1685}.

\bibitem{abn}J.P. Ader, F. Biet and Y. Noirot, {\em Super w-models and
their anomalies : a geometrical approach}, preprint CPTMB/PT/95-1.


\end{thebibliography}
\end{document}